\documentclass[useAMS,usenatbib]{mn2e}
\usepackage{graphicx}
\usepackage{epsfig}
\usepackage{subfigure}
\usepackage{color}
\usepackage{xcolor}
\usepackage{multirow}
\usepackage{threeparttable}
\usepackage{natbib}
\usepackage{amsmath,ulem,upgreek,amssymb}
\usepackage[colorlinks=true,linkcolor=blue,citecolor=blue]{hyperref} 

\newcommand{\hiz}{high-$z$ }

\newcommand{\zphot}{$z_{\mathrm{ph}~}$}
\newcommand{\Ang}{\text{\AA}}

\definecolor{bluet}{HTML}{4169e1}

\title[RXC J2248  $z\sim6$ quintuply lensed candidate]{CLASH: $z\sim6$ young galaxy candidate quintuply lensed by the frontier field cluster RXC J2248.7-4431}
\author[A. Monna et al.]
{\parbox{\textwidth}{\Large A. Monna$^{1,2}$\thanks{E-mail:
amonna@usm.uni-muenchen.de},
S. Seitz$^{1,2}$,
N. Greisel$^{1,2}$,
T. Eichner$^{1,2}$,
N. Drory$^{2,3}$,
M. Postman$^{4}$,
A. Zitrin$^{5,24}$,
D. Coe$^{4}$,
A. Halkola$^{}$,
S. H. Suyu$^{6}$,
C. Grillo$^{7}$,
P. Rosati$^{8}$,
D. Lemze$^{9}$,
I. Balestra$^{2}$,
J. Snigula$^{1,2}$,
L. Bradley$^{4}$,
K. Umetsu$^{6}$,
A. Koekemoer$^{4}$,
U. Kuchner$^{23}$,
L. Moustakas$^{20}$,
M. Bartelmann$^{5}$,
N. Ben\'itez$^{10}$,
R. Bouwens$^{11}$,
T. Broadhurst$^{12}$,
M. Donahue$^{13}$,
H. Ford$^{9}$,
O. Host$^{7}$,
L. Infante$^{14}$,
Y. Jimenez-Teja$^{10}$,
S. Jouvel$^{15,16}$,
D. Kelson$^{17}$,
O. Lahav$^{15}$, 
E. Medezinski$^{9}$,
P. Melchior$^{18}$,
M. Meneghetti$^{19}$,
J. Merten$^{20}$,
A. Molino$^{10}$,
J. Moustakas$^{21}$,
M. Nonino$^{22}$,
W. Zheng$^{9}$}\vspace{0.4cm}\\
\parbox{\textwidth}{$^{1}$University Observatory Munich, Scheinerstrasse 1, 81679 Munich, Germany\\
$^{2}$Max Planck Institute for Extraterrestrial Physics, Giessenbachstrasse, 85748 Garching, Germany\\
$^{3}$Instituto de Astronom\'ia, Universidad Nacional Aut\'onoma de M\'exico, Avenida Universidad 3000, 04510, D.F. Mexico\\
$^{4}$Space Telescope Science Institute, 3700 San Martin Drive, Baltimore, MD 21208, USA\\
$^{5}$Institut f\"ur Theoretische Astrophysik, ZAH, Albert-Ueberle-Stra e 2, 69120 Heidelberg, Germany\\
$^{6}$Institute of Astronomy and Astrophysics, Academia Sinica, P.O. Box 23-141, Taipei 10617, Taiwan\\
$^{7}$Dark Cosmology Centre, Niels Bohr Institute, University of Copenhagen, Juliane Maries Vej 30, 2100 Copenhagen, Denmark\\
$^{8}$ESO-European Southern Observatory, D-85748 Garching bei M\"unchen, Germany\\
$^{9}$Department of Physics and Astronomy, The Johns Hopkins University, 3400 North Charles Street, Baltimore, MD 21218, USA\\
$^{10}$Instituto de Astrof\'isica de Andaluc\'ia (CSIC), Camino Bajo de Huetor 24, Granada 18008, Spain\\
$^{11}$Leiden Observatory, Leiden University, P. O. Box 9513,2300 RA Leiden, The Netherlands\\
$^{12}$Department of Theoretical Physics, University of the Basque Country, P. O. Box 644, 48080 Bilbao, Spain\\
$^{13}$Department of Physics and Astronomy, Michigan State University, East Lansing, MI 48824, USA\\
$^{14}$Departamento de Astrono\'ia y Astrof\'isica, Pontificia Universidad Cat\'olica de Chile, V. Mackenna 4860, Santiago 22, Chile\\
$^{15}$Department of Physics \& Astronomy, University College London, Gower Street, London WCIE 6 BT, UK\\
$^{16}$ Institut de Cincies de l'Espai (IEEC-CSIC), Bellaterra (Barcelona), Spain\\
$^{17}$Observatories of the Carnegie Institution of Washington, Pasadena, CA 91 101, USA\\
$^{18}$Center for Cosmology and Astro-Particle Physics; The Ohio State University, 191 W. Woodruff Ave., Columbus, Ohio 43210, USA\\
$^{19}$INAF-Astronomical Observatory of Bologna, Via Ranzani 1, 40127 Bologna, Italy\\
$^{20}$Jet Propulsion Laboratory, California Institute of Technology, MS 169-327, Pasadena, CA 91109, USA\\
$^{21}$Siena College, 515 Loudon Road, Loudonville, NY 12211,USA\\
$^{22}$INAF-Osservatorio Astronomico di Trieste, via G.B. Tiepolo 11, 40131 Trieste, Italy\\
$^{23}$University of Vienna, Department of Astrophysics, T\"urkenschanzstr.17, 1180 Wien, Austria\\
$^{24}$Hubble fellow; Cahill Center for Astronomy and Astrophysics, California Institute of Technology,MS 249-17, Pasadena, CA91125,USA\\
}}

\begin{document}
\date{}
\pagerange{\pageref{firstpage}--\pageref{lastpage}} \pubyear{2013}
\maketitle
\label{firstpage}
\begin{abstract}
We present a quintuply lensed $z\sim6$ candidate discovered in the field of the galaxy cluster RXC J2248.7-4431 ($z\sim0.348$) targeted within the Cluster Lensing and Supernova survey with Hubble (CLASH) and selected in the deep HST Frontier Fields survey.
Thanks to the CLASH 16-band HST imaging, we identify the quintuply lensed $z\sim6$ candidate as an optical dropout in the inner region of the cluster,  the brightest image having $\rm mag_{AB}=24.8\pm0.1$ in the f105w filter. We perform a detailed photometric analysis to verify its \hiz and lensed nature. We get as photometric redshift \zphot$\sim5.9$, and given the extended nature and NIR colours of the lensed images, we rule out low-z early type and galactic star contaminants.
We perform a strong lensing analysis of the cluster, using 13 families of multiple lensed images identified in the HST images. Our final best model predicts the \hiz quintuply lensed system with a position accuracy of 0.8$\arcsec$. The magnifications of the five images are between 2.2 and 8.3, which leads to a delensed UV luminosity of $L_{1600}\sim0.5L_{1600}^*$ at $z=6$.  We also estimate the UV slope from the observed NIR colours, finding a steep $\beta=-2.89\pm0.38$. We use singular and composite stellar population SEDs to fit the photometry of the \hiz candidate, and we conclude that it is a young (age $<300$\,Myr) galaxy with mass of $M\sim10^8M_{\odot}$, subsolar metallicity ($Z<0.2Z_\odot$) and low dust content $(A_V\sim0.2-0.4)$.
\end{abstract}


\section{Introduction}
\label{sec:Introduction}
The Cluster Lensing And Supernovae survey with Hubble (CLASH, \citealt{Postman2012a}), is a 524-orbit multicycle treasury Hubble Space Telescope (HST) program, which is targeting 25 clusters of galaxies with the main goal to measure accurate cluster mass profiles by combining space-based strong lensing  analyses and ground-based weak lensing analyses (see \citealt{Umetsu2012}, \citealt{Coe2012}, \citealt{Medezinski2013}). 
 \begin{figure*}
 \centering
 \includegraphics[width=16cm]{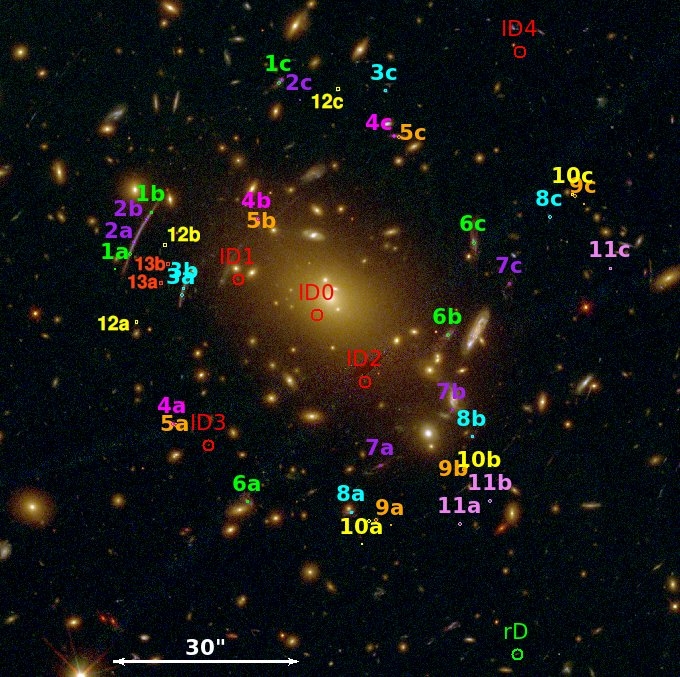}
 \caption{\small $\rm2'\times2'$ colour composite image of RXC J2248.7-4431 core, from the HST 16-band of the CLASH survey: Blue=F435w+F475w; Green=F606w+F625w+F775w+F814w+F850lp; Red=F105w+F110w+F140w+F160w. We label the multiple lensed systems used in the strong lensing analysis (see Sec.~\ref{sec:lensing_evid}). 
 The systems 12, 13 and the lensed image 11.c  are new systems recently identified in the model. We verified that including these systems in the model is not changing significantly the final best model nor the errors on the model parameters (see Sec.~\ref{sec:lensing_evid} for more details). The red circles (ID0-4) label the quintuply lensed dropout candidate at $z\sim6$. The central image ID0 is selected after removing the BCG light (see Sec.~\ref{sec:central_image}). With rD we label the other r-dropout candidate at z$\sim5$ identified through photometric selection (see Sec.~\ref{sec:procedure}).}
         \label{fig:rgb}
 \end{figure*}
In addition, taking advantage that galaxy clusters act as gravitational lenses, CLASH also aims to identify and investigate magnified galaxies which contribute to reionize the intergalactic medium in the early Universe at redshift $z\gtrsim6$.
Up to date several lensed 
galaxies at $z>5$ are identified in galaxy clusters fields, e.g.,  a galaxy at $z\sim6$ lensed into 3 images in A383 \citep{Richard2011} (spectroscopically confirmed), a lensed galaxy at $z\sim6.56$ in A370 \citep{Hu2002} (spectroscopically confirmed),  seven lensed galaxies at $z\sim7$ in A1703 \citep{Bradley2012a} (one of which spectroscopically confirmed), a quadruply lensed galaxy at $z\sim6.2$ in the field of MACS0329 \citep{Zitrin2012}, a triply lensed galaxy at z$\sim7$ in A2218 \citep{Kneib2004}, a highly magnified  galaxy at $z\sim9.6$ \citep{Zheng2012} in MACS1149 and a triply lensed galaxy at $z\sim11$ in MACS0647 \citep{Coe2012}. The last two sources are the highest redshift lensed candidates detected to date, and were discovered within the CLASH survey. A large sample of single and multiply lensed \hiz candidate galaxies identified in the CLASH survey is described and analysed in \citet{Bradley2013}.\\ In this paper we present a candidate quintuply lensed galaxy at $z\sim6$ discovered in the field of the galaxy cluster RXC J2248.7-4431 (here after RXC J2248) that we identify thanks to the wide CLASH photometric dataset (see Fig.~\ref{fig:rgb}).\\

RXC J2248 is a very massive galaxy cluster ($M_{200}>2.5\cdot10^{15}M_\odot$) at $z=0.348$ \citep{2009A&A...499..357G}. 
It is part of the ROSAT-ESO Flux-Limited X-ray (REFLEX) survey galaxy cluster sample, 
and it is the second brightest cluster in this survey, with a luminosity of $L\sim3\times10^{45}$\,erg\,s$^{-1}$ in the rest frame 0.1-2.4\,KeV interval \citep{2009A&A...499..357G}.
\citet{gomez2012} 
confirmed 51 cluster members  with spectroscopic observations taken with the Gemini Multi-Objects Spectrograph (GMOS), from which they estimate the mean effective velocity dispersion and redshift of the cluster to be $\sigma=1660^{+230}_{-150}$\,km/s and $z=0.3461^{+0.0010}_{-0.0011}$. 
The galaxy density, obtained using spectroscopically confirmed  and  candidate (selected from the colour magnitude diagram) cluster members, combined with the X-ray isophotes, reveals  several substructures and an elongation in NE-SW direction.
The X-ray analysis  shows that the X-ray peak emission is shifted by $37''\pm9''$ with respect to the galaxy density peak. 
The combination of optical, spectroscopic and X-ray analysis thus reveals traces of recent merging activities.
 \citet{Gruen2013} performed a multi-wavelength weak lensing analysis of this cluster using deep UBVRIZ imaging data  from the ESO-2.2m Wide Field Imager, with a total integration time of about 50 hours. They accurately  measured the 2D mass distribution and mass profile for scales larger than 500 kpc, out to 4 Mpc. Their mass map shows that the main dark matter halo is centred on the  brightest cluster galaxy (BCG), and that dark matter substructures in and around the cluster coincide with galaxy density substructures, dominated by red galaxies.  
 Fitting the dark halo with an NFW profile, they obtain the values $ c_{200m}=2.6^{+1.5}_{-1.0}$ and $M_{200m}=33.1^{+9.6}_{-6.8}\cdot10^{14} M_{\odot}$, in agreement with previous X-ray and SZ estimates. Due to its strong lensing strength, RXC J2248 is selected in the sample of 6 galaxy clusters of the deep Frontier Fields (FF) HST survey\footnote{http://www.stsci.edu/hst/campaigns/frontier-fields/}, which aims to map the cluster dark matter to the highest precision and derive magnification maps to investigate the high redshift Universe.\\ 
 
Our paper is organised as follows. In Section~\ref{sec:Dataset} we present the photometric data and redshifts.  In Section~\ref{sec:procedure} we provide an overview of the dropout selection procedure we use and present the $z>5$ candidates selected in the field of RXC J2248, including the multiply lensed $z\sim6$ galaxy. In Section~\ref{sec:photometric_evid} we discuss the photometric evidence supporting the lensed \hiz nature of our candidate. In Section~\ref{sec:lensing_evid} we provide the strong lensing analysis of the cluster, and we show that the lensing model predicts our candidate to be at high redshift. In Section~\ref{sec:central_image} we present the fifth (central) image of the system, identified thanks to the lensing model and detected after subtracting the BCG light from the images. In Section~\ref{sec:Properties} we analyse the physical properties of the $z\sim6$ candidate and compare it with other similar systems analysed in the CLASH survey. In Section~\ref{sec:conclusions} we 
present our summary and conclusions.\\
In this paper we adopt the concordance $\Lambda$CDM cosmology with $h=0.7$, $\Omega_{m0}=0.3$ and $\Omega_{\Lambda0}=0.7$.   
\begin{table}
\caption{RXC J2248 CLASH Dataset summary: column (1) filters included in the survey, column (2) total observation time in seconds, column (3) HST instrument, column (4) the $5\sigma$ magnitude depth within $0.6\arcsec $.}
\centering
\footnotesize
\begin{tabular}{|c|c|c|c|}
\hline
\hline
Filter&Total time (s)&Instrument&5$\sigma$ Depth\\
\hline
f225w &    7148 & WFC3/UVIS & 25.43\\
f275w &    7274 & WFC3/UVIS & 25.44 \\
f336w &    4718 & WFC3/UVIS & 25.74 \\
f390w &    4740 & WFC3/UVIS & 26.47 \\
f435w &    4102 & ACS/WFC   &   26.35  \\    
f475w &    2064 & ACS/WFC   &   26.77    \\  
f606w &    3976 & ACS/WFC   &   27.07      \\
f625w &    4128 & ACS/WFC   &   26.60      \\
f775w &    4058 & ACS/WFC   &   26.30      \\
f814w &    8136 & ACS/WFC   &   26.94      \\
f850lp &    6164 & ACS/WFC   &   25.84     \\
f105w &    2815 & WFC3/IR   &   26.86 \\
f110w &    2515 & WFC3/IR   &   26.87 \\
f125w &    1509 & WFC3/IR   &   26.88   \\
f140w &    2312 & WFC3/IR   &   26.93   \\
f160w &    3520 & WFC3/IR   &   26.96   \\
\hline                  
\end{tabular}
\label{tab:phot}
\end{table}

\section{Photometric Dataset}
\label{sec:Dataset}
RXC J2248.7-4431 was observed between September 2012 and October 2012 in 16 filters covering the UV, optical and NIR range with the HST Advanced Camera for Surveys (ACS) and the  HST Wide Field Camera 3 (WFC3) UVIS and IR cameras.
See Tab.~\ref{tab:phot} for the complete list of the filters.
\begin{figure}
\centering
\includegraphics[width=8.5cm]{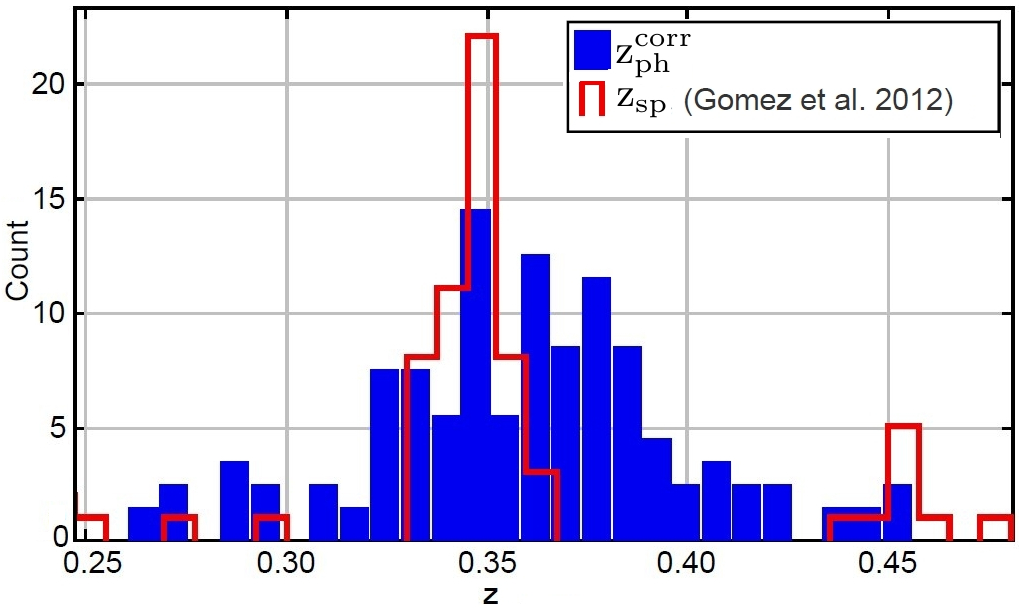}
\caption{\small Redshift distribution of sources extracted in the  WFC3IR FOV of RXC J2248.
In blue we plot the distribution of the corrected photometric redshifts $z_{\rm ph}^{\rm corr}$ computed with \texttt{LePhare}  for sources with $\rm f775w\_mag\_best<23$, within the range $z_{\rm ph}=[0.25,0.45]$.
In red we plot the distribution of spectroscopic redshift from \citet{gomez2012}.
In the $z_{\rm ph}$ distribution a peak is identified in correspondence of the spectroscopic $z\rm_{cl}$.}
        \label{fig:z_cl_histo}
\end{figure}
The HST dataset is composed of mosaic drizzled science images and respective weight images produced with the \texttt{Mosaicdrizzle} pipeline \citep[see][]{2011ApJS..197...36K}. They have pixel scale of 65mas$/$pixel and cover a field of view (FOV) of $\sim 3.4'\times3.4'$ in the ACS images and $\sim2'\times2'$ in the WFC3IR images. Using \texttt{SExtractor} 2.5.0 \citep{1996A&AS..117..393B}, we obtain a multiband photometric catalogue of galaxy fluxes extracted within  $0.6\arcsec$(9~pixels) diameter aperture.
We run \texttt{SExtractor} in dual image mode, using as detection image the weighted sum of all the WFC3IR images (see Sec.~\ref{sec:procedure} for more details about the choice of the detection image). 
We independently estimate the true aperture photometric uncertainties in each filter and compare them with the formal \texttt{SExtractor} uncertainties.  We measure the  real photometric errors for each detected source using the aperture photometry and sky background\footnote{using the formula $\rm N=\sqrt{S + n_{pix}\times B}$, where N is the noise, S is the signal, $\rm n_{pix}$ is the number of pixels in the aperture and B is the background per pixel.
%$\rm rms=\sqrt{\frac{flux{_ap}}{gain}+(n_{pix}\times R_{backgr}))^2}$, where $\rm flux_{ap}$ is the flux extracted in the 9 pixel aperture, $\rm n_{pix}$  is the number of pixel in the aperture, and $\rm R_{backgr}$ is the local sky background.
} and we find that the \texttt{SExtractor} errors are underestimated by a mean factor of 4 in our dataset. In the rest of the paper we use photometric uncertainties corrected by this factor.
Moreover we compute the flux detection limit in each band by measuring the fluxes within $3000$ random apertures of $0.6''$ diameter within the image FOV. We then fit a gaussian to the flux measurement histogram, estimate the $1\sigma$ width and obtain the 5$\sigma$ limiting magnitudes. The results are given in  Tab.~\ref{tab:phot}.
\begin{figure}
\centering
\includegraphics[width=8cm]{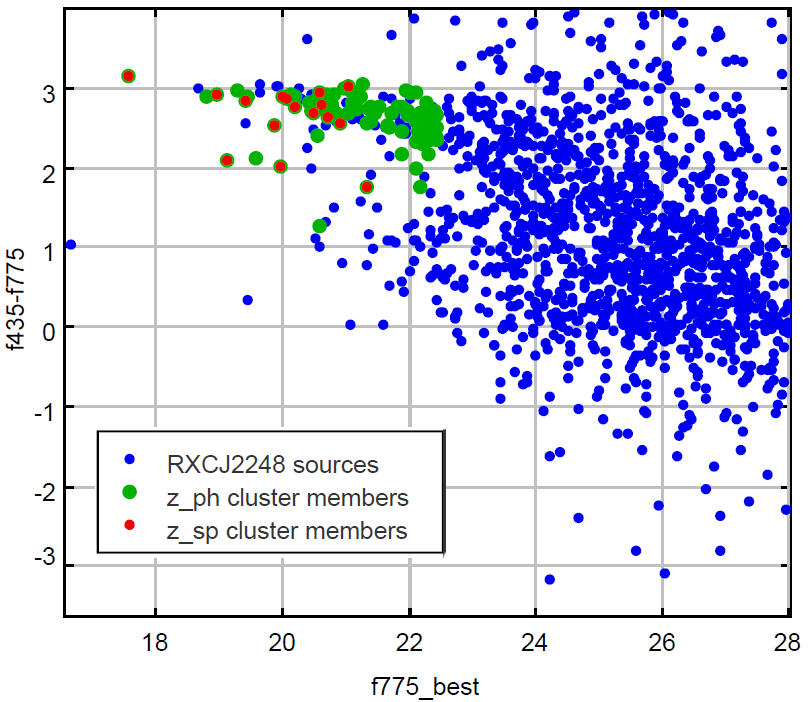}
\caption{\small Colour magnitude diagram for sources extracted in the  WFC3IR FOV of RXC J2248.
We plot the colour from the aperture magnitudes f435w and f775w, versus the \texttt{SExtractor} mag$\_$best in the f775w filter.
Blue circles are all the sources extracted in the  WFC3IR FOV;
Red circles are the spectroscopic confirmed cluster members from \citet{gomez2012};
Green circles are cluster member candidates with $z\rm_{ph}\in[0.25,0.45]$ and $\rm f775w\_mag\_best<22.5$.
We select 86 bright cluster member candidates within the $2\arcmin\times2\arcmin$ WFC3IR FOV, which almost all lie on the red sequence of the cluster.}  
        \label{fig:col_mag}
\end{figure}

\subsection{Cluster members: photometric and spectroscopic redshifts}
\label{sec:photo_z}
We use the spectral energy distribution (SED) fitting code \texttt{LePhare}\footnote{http://www.cfht.hawaii.edu/~arnouts/lephare.html} \citep{1999MNRAS.310..540A,2006A&A...457..841I} to compute the photometric redshifts of  galaxies to use for the selection of cluster members. The package computes photometric redshifts through 
a standard $\chi^2$ fitting method to fit the observed fluxes with template spectra,  where the $\chi^2$ is defined as \citep{2006A&A...457..841I}
\begin{equation}
\chi^2(z,T,A)=\sum_{f=1}^{N_F}\left(\frac{F^{f}_{\mathrm{obs}}-A\times F^{f}_{\mathrm{pred}}(z,T)}{\sigma_{\mathrm{obs}}^f}\right)^2
\end{equation}

where the index $f$ refers to the filters of the dataset, $N_f$ is the number of filters, $F^{f}_{\rm obs}$ is the  observed flux, $\sigma_{\mathrm{obs}}^f$ is its error, $F^{f}_{\mathrm{pred}}(z,T)$ is the estimated flux for the redshift $z$ and the spectral type $T$, and finally $A$ is a normalisation factor. 
The photometric redshift is computed by searching for the solution $(z,T,A)$ for which the $\chi^2$ is minimal.\\
Galaxy, QSO and star SED templates can be used in the templates set; moreover, extinction laws, emission lines and constraints on the range of redshift, age, absolute magnitude and other physical parameters can be included in the fitting procedure \citep[see][]{1999MNRAS.310..540A,2006A&A...457..841I}.\\ 
The output files give the photometric redshift values for the best fits each for the star, QSO and galaxy templates: in particular for the galaxy solutions, \texttt{LePhare} computes the redshift probability distribution function (PDF($z$)) and gives also the secondary solution from the PDF($z$), if available.\\
We adopt as galaxy template set the COSMOS libraries \citep{2009ApJ...690.1236I}, which include 31 templates describing SEDs of ellipticals, spirals and starburst galaxies.
To take into account the extinction due to the interstellar medium (ISM) we apply the Calzetti extinction law \citep{2000ApJ...533..682C} to the starburst templates, and the SMC Prevot law \citep{Prevot1984} to the Sc and Sd templates.
We also include emission lines, and add prior information on the redshift distribution: 
for that the code uses an $N(z)$ prior by type computed from the VVDS survey \citep[see][]{2006A&A...457..841I,2000ApJ...536..571B}.
As QSOs template set we use the SWIRE library \citep{Polletta2006, Hatziminaoglou2005,Gregg2002} including type 1, type 2,  Seyfert 1.8 and Seyfert 2 AGN templates. Finally as stellar templates we adopt the Pickles stellar library \citep{Pickles1998}, which include all the normal spectral types, plus metal-poor F-K dwarves and G-K giants.\\ 
\citet{Greisel2013} carried out a systematic analysis of luminous red galaxies from the Sloan Digital Sky Survey (SDSS) with $0<z<0.5$, which shows that local SED templates do not match within appropriate level the observed colours for galaxies at different redshifts, due to the evolution  of the SEDs with redshift. Thus we expect that, using \texttt{LePhare} libraries to estimate the cluster members photo-z, we predict different colours from what we actually observe in some filters. Indeed when we estimate the photometric redshift for the sources extracted in our images, the cluster appears to be at $z\sim0.4$ from the \zphot histogram. One way to account for the template mismatch of red SEDs at the $z_{\rm cl}=0.348$ is to apply some offsets correction to our photometry.
To estimate these offsets we use the sample of spectroscopically confirmed cluster members. 
In \texttt{LePhare} they are estimated 
 through a colour adaptive method which finds the template that best fits the observed photometry given the fixed $z_{\mathrm{sp}}$ of the galaxies, and minimises the offset between the observed and predicted magnitudes in each filter. We use 16 cluster members from the spectroscopic sample \citep{gomez2012}, which are in our WFC3IR FOV, to compute the magnitudes offsets in our filters, and apply  these offsets to the observed photometry 
to compute the photometric redshift for the cluster member galaxies.
The offsets we get are not significant ($<0.03$ mag) in the filters from f606w to f160w, while they are higher (between 0.1 and 0.36 mag ) in the blue filters (from f475w to f225w) showing that the adopted templates do not describe the early type restframe SEDs well for wavelengths smaller than  3500\AA. 
In Fig.~\ref{fig:z_cl_histo} we plot the histogram of the  offset corrected $z\rm_{ph}^{corr}$ for the bright ($\rm f775w\_mag\_best<23$) galaxies extracted in our images, which shows a peak at the cluster redshift $z\rm_{cl}\sim0.35$ when using the estimated magnitude offsets.  
We select 80 bright cluster member candidates with $z\rm_{ph}^{corr}$ in the range $[0.25,0.45]$ and $\rm f775w\_mag\_best<23$ best fitted with early type templates, all of them lying on the red sequence in the colour$-$magnitude diagram (see Fig.~\ref{fig:col_mag}).
We use this  cluster members sample as the galaxy component in performing the strong lensing mass model of the cluster (see Section~\ref{sec:lensing_evid}).

\section[]{High-$z$ dropouts: photometric selection}
\label{sec:procedure}
\begin{figure}
   \includegraphics[width=8cm]{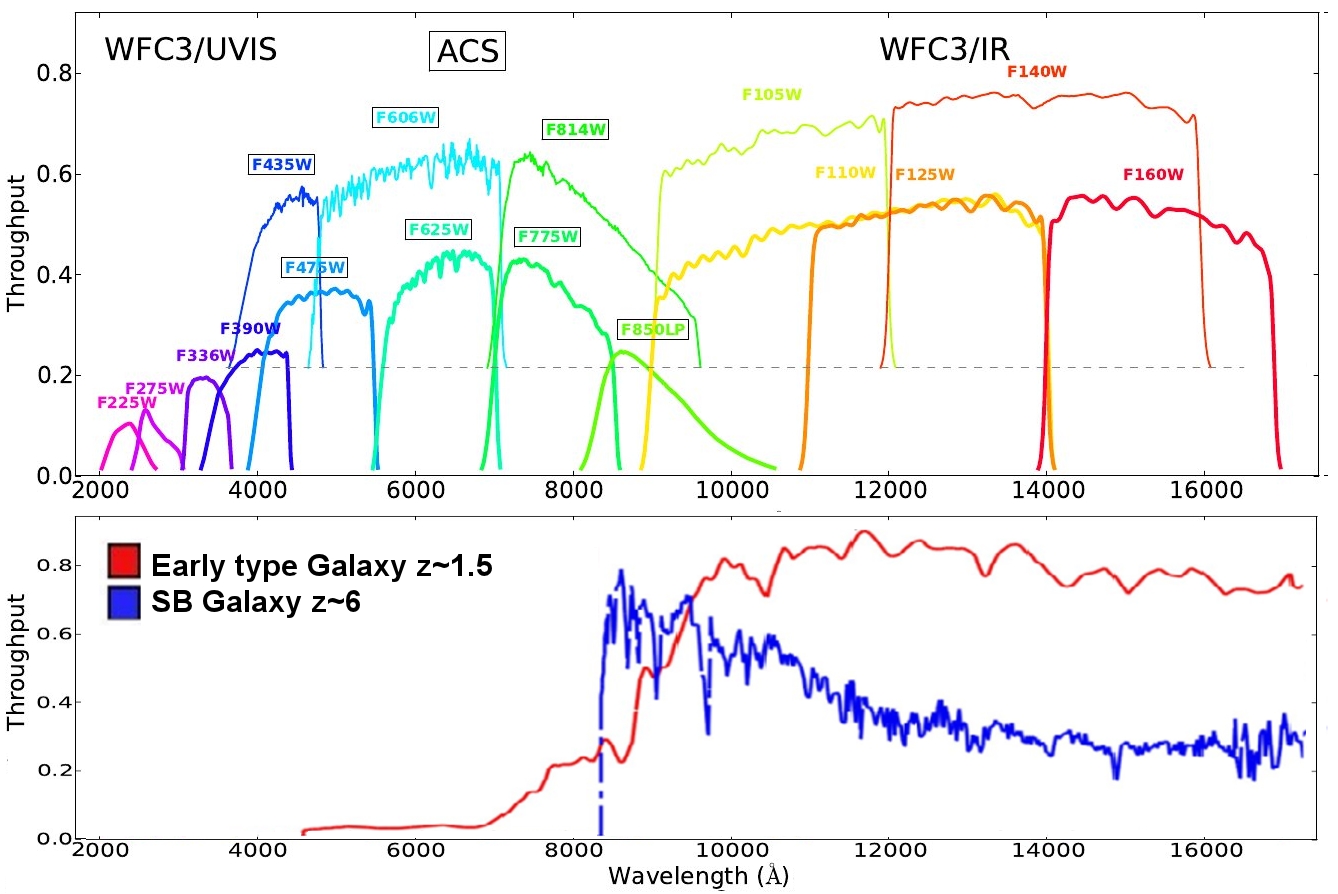}
   \caption{Upper panel: transmission curves of the HST/ACS and HST/WFC3 filters used in the CLASH survey. Lower panel: Comparison of a star-forming SED redshifted to $z=6$ (in blue) and an early-type galaxy at $z=1.5$ \citep[from the COSMOS library,][]{2009ApJ...690.1236I}. In both cases the strong spectral break falls in the same NIR range, and both these kind of sources appear as optical dropouts. At $\lambda$ redder than the break the SEDs differ, thus the NIR photometry can be used to discriminate between high- and low-$z$ dropouts.}%
      \label{fig:filters} 
 \end{figure}

Star forming galaxies at high redshift are known as Lyman Break Galaxies \citep[LBGs, see][]{Steidel1996c,Dunlop2013rev}, because of the strong break in the SED at the Lyman$-\alpha$ wavelength ($\rm \lambda_{Ly\alpha}=1216$\AA$ $ restframe) due to the intergalactic medium (IGM) absorption: the UV flux emitted at $\rm lambda < \lambda_{Ly\alpha}$ is absorbed by the neutral hydrogen (HI) of the inter galactic medium (IGM)  along the path travelled by the light, causing a peculiar break in the observed SED of these sources at the redshifted $\rm\lambda_{Ly\alpha}$.
This effect is known as ``Gunn-Peterson trough'': the optical depth $\tau_{GP}$ of Lyman-$\alpha$ photons is directly proportional to the HI density in the IGM. At high redshift ($z>5$), a tiny amount of HI (for example a fraction $\rm x_{HI}\sim10^{-4}$) leads to a complete absorption of UV photons with $\rm\lambda < \lambda_{Ly\alpha}$ \citep[see][]{2006ARA&A..44..415F}.    
Thus candidate galaxies at redshift $z>5$ can be identified in photometric datasets using the so called 'dropout technique' \citep{Steidel1996a,Steidel1996b}, which aims to identify the Lyman-$\alpha$ break in galaxy SEDs through colour analyses.
In particular, for galaxies at $z>5$ the Lyman break is redshifted to the near infra-red spectral range, and these sources are expected to be not detected in the UV and optical filters (see Fig.~\ref{fig:filters}).\\
The optical dropout nature cannot be used alone to identify \hiz galaxies, since other galaxies mimic the same photometric behaviour: these are early-type galaxies at lower ($z\sim1$) redshift, for which the dropout is due to the restframe  4000\AA\ break, caused by the stellar photospheric opacity shortward of 4000 \AA. In this case the flux blueward this break can be fainter than the detection limit,  leading to none detection in the optical filters (see \citealt{2006ApJ...645...44K}).
To discriminate between \hiz star-forming and low-$z$ early-type galaxies we need to measure the colours at wavelength redder than the observed break: indeed comparing the SEDs of star-forming galaxies at redshift $z>5$ and  early-type galaxies at $z\sim1.5$, the colours at wavelengths larger than the dropout filter are expected to be blue for the former and red for the latter, see Fig.~\ref{fig:filters}.\\

\begin{figure}
\centering
\includegraphics[width=8.cm]{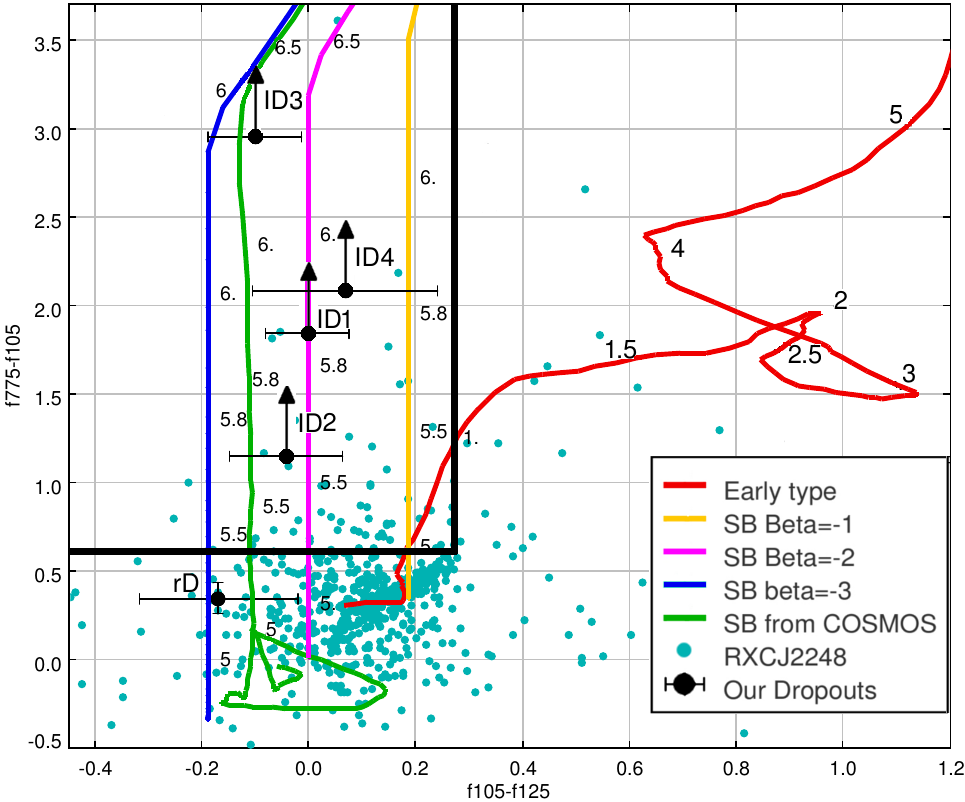}
\caption{\small Optical and NIR colours diagram for the sources in the field of the CLASH cluster RXC J2248.
In light blue we plot all the sources extracted in the WFC3IR FOV, in black our \hiz dropout candidates. 
We also plot the colour-redshift tracks for the SB galaxies that we generated with UV continuum slope $\beta=-1,-2,-3$ (in yellow, magenta and blue respectively), for a SB  (in light green)  and an early type (in red) templates from the COSMOS library (Ilbert et al. 2009).
The black lines enclose the diagram region satisfying the colour criteria for z$>5$ candidates from Eq.~\ref{eq:color_criteria}.   }
\label{fig:col_col_diagr}
\end{figure}

As described in Sec.~\ref{sec:Dataset}, we build the multiband photometric catalogue of fluxes in $0.6''$ diameter apertures. 
We use the weighted sum of all the WFC3IR images as the detection frame since our aim is to identify sources at $z\ga5$ % 
which  are expected to be detected in the NIR filters.  
Moreover, since \hiz  candidates are expected to have small size (e.g.,  the galaxy half light radius at $z=6$ is expected to be $r_{\rm hl}\sim1\rm kpc$, see \citealt{Bouwens2004}), we use a $\rm DETECTION\_MIN\_AREA$ of 9 pixels as the minimum number of pixels above the threshold, we use the value of 1 as absolute detection treshold and we apply the gaussian Sextractor filter \texttt{gauss$\_2.0\_5$x$5$.conv} in the detection.  
Our catalogue of sources extracted in the WFC3IR$\_$total image counts 7767 objects: of course, given the small $\rm DETECTION\_MIN\_AREA$ and the low detection threshold adopted for the extraction, several of these sources are spurious detections due to background fluctuations, but most of them will be automatically removed from our sample through our dropout selection procedure.\\  
To select optical dropout candidates at $z\ga5$, we compute photometric redshifts with \texttt{LePhare} (see Sec.~\ref{sec:photo_z}) using the multiband catalogue of aperture fluxes, and using the $1\sigma$ detection limit as upper limit when sources are not detected in a filter.
Then we select all the sources with first or secondary galaxy \zphot solution higher then 5.\\
We use the starbust (SB) colour-redshift tracks to define the colour criteria that sources need to satisfy as \hiz candidates. In the UV continuum range ($1250-2600$ \AA), SB SEDs follow the law $\rm F_\lambda \propto \lambda^\beta$ (see Calzetti et al. 1994), with an expected mean $\beta=-2$ at $z\sim6$ (see appendix in \citealt{Bouwens2012}).
Thus we generate SB SEDs with UV continuum slope $\beta=-1,-2,-3$, and compute the colour expected for these templates in the CLASH filters as a function of the redshift in the range $z \in [0,12]$. In Fig.~\ref{fig:col_col_diagr} we plot the SB redshift tracks with $\beta=-1,-2,-3$ in the optical and NIR colours diagram, together with the tracks of an early type galaxy (in red) and a SB galaxy (in green) from the COSMOS library \citep{2009ApJ...690.1236I}.    
Referring to SB tracks 
we find that the colours for star forming galaxies at $z>5$ are expected to be (see Fig.~\ref{fig:col_col_diagr}):
\begin{equation}
\rm (i_{775}-Y_{105} > 0.6) \wedge (Y_{105}-J_{125} < 0.3).
\label{eq:color_criteria}
\end{equation}
We check the optical and NIR colours of our \hiz candidates (all the galaxies with primary or secondary \zphot solution) and reduce the sample to objects matching these colour criteria. Moreover,  we require that all the dropouts have detection fainter than the $5\sigma$ detection limit in the filters bluer than the f606w, and detection higher than $5\sigma$ in one or more of the redder filters. To remove stellar contaminants (galactic M, L and T dwarf stars) and reduce the sample to secure extragalactic sources, we reduce our \hiz dropouts candidates to the extended sources. 
Finally, we perform a visual check to remove spurious detections, like objects at the images edges, or stellar sparks.\\%, or

\begin{figure*}
 \centering
  \includegraphics[width=18cm]{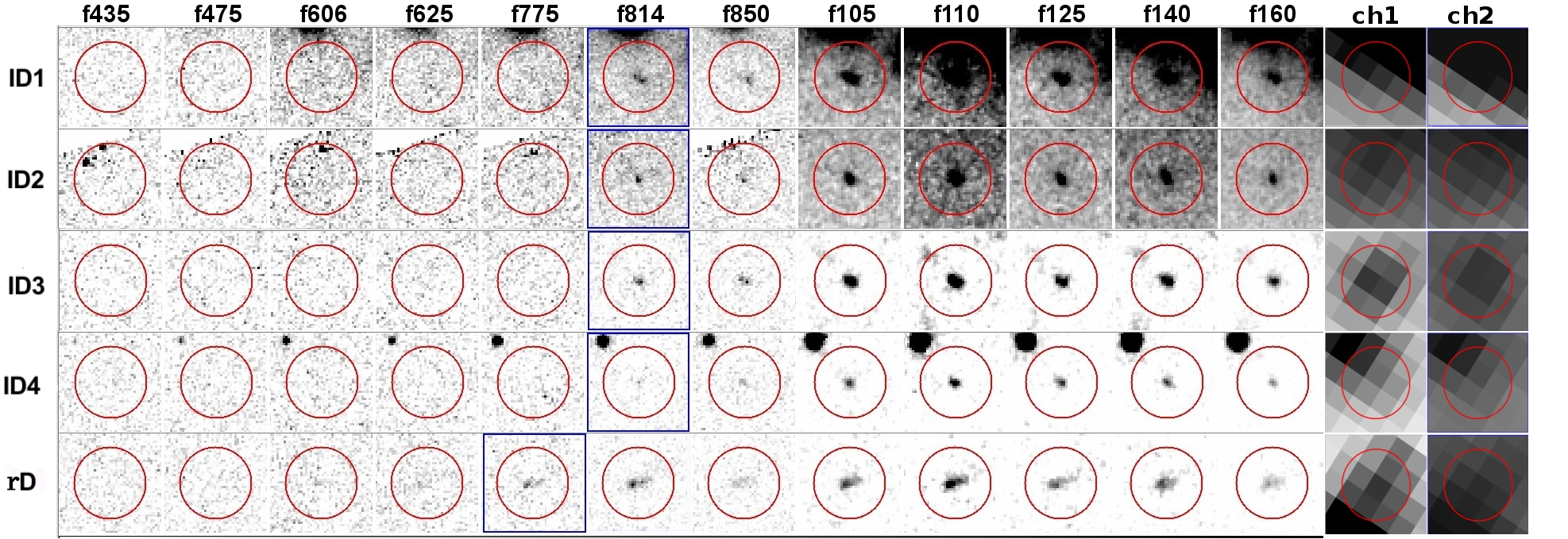}
 \caption{ Postages of the dropouts candidate at z$>5$ in the HST filters from the f435 to the IRAC/45\,$\upmu$m. The first four objects (ID$1-4$) are the candidate z$\sim6$ multiple lensed system: for this system the first detection band is the f814w, while in the bluer filters there is not significant detection (flux $<5\sigma$). The last object (rD) is a r-band dropout, with first significant detection $5\sigma$ in the f775w filter. The red circles have diameters of $\rm 2\arcsec$, i.e. $\rm \sim3$ times the aperture diameter  we use to extract the photometry. }
     \label{fig:dropouts}
 \end{figure*}

Our final sample yields 5 \hiz candidate dropouts. All of them are extended sources with detection $>5\sigma$ in the NIR filters (see Tab.~\ref{tab:mags} for  the aperture photometry in the CLASH HST filters).\\
 We supplement our photometric dataset with mosaic mid-IR imaging data from Spitzer
 obtained with the Infrared Array Camera (IRAC) in  channel1 (3.6\,$\upmu$m) and channel2 (4.5\,$\upmu$m), from the survey "Use of massive clusters as cosmological lenses"(PI: G. Rieke, program ID 83). We analyse these images and find that three of our \hiz candidates (ID2, ID3 and rD) are detected in both IRAC channels, even if they are very faint. 
 The photometric contamination by the bright close by cluster members is very high in these mid-IR images, in particular if compared to the optical and NIR HST images. 
 This contamination, also aggravated by the  Spitzer/IRAC PSF being larger than that of the HST,
 does not allow for estimation of robust uncertainties on the extracted photometry for our candidates. Thus, for these candidates we extract the mid-IR photometry within aperture of $1.2\arcsec$, and we use the resulting values as upper limits in our following SED fits. 
In Fig.~\ref{fig:dropouts}, we present the  postage stamp images from the f435w to the IRAC/45$\upmu$m filter, and in Tab.\ref{tab:mags} we provide the aperture photometry in the filters from f625 to IRAC/45$\upmu$m. The four objects ID1-4 have the lowest wavelength detection in the $i$ band: ID1-3 are first detected in the f814w filter with magnitude $>5\sigma$ and ID4 is detected with magnitude $>3\sigma$ in the f814w filter and $>5\sigma$ in the f850lp filter. 
The \hiz candidate rD is a dropout in the $r$ band: it is not detected in the filter bluer than f625w, in which is detected at the $3\sigma$ limit, and it has detection $>5\sigma$ in the f775w and redder filters. For this candidate we get z$_{\rm phot}\sim5$, in agreement with the $>5\sigma$ detection in f775w and with the fainter detection in the f625w filter.
In Tab.~\ref{tab:dropouts} we summarise the photometric redshift results from \texttt{LePhare}: all the candidates have only photometric redshift solution at high redshift, \zphot$>5$, for galaxy templates. 
ID1-4 have all similar photometric redshifts of $z\sim5.9$ with a well defined peak in the PDF(z), see Sec.~\ref{sec:photometric_evid}, and their positions on the cluster FOV suggest that they can be a system of multiple images of the same source lensed by the cluster.
%It has primary solution for the photometric redshift at $\rm z\sim5$. 
 In the following sections we will focus on the system ID1-4 to verify its \hiz multiple lensed nature through photometric and lensing analyses.
\begin{table*}
\caption{Summary of the HST photometry extracted within apertures of  $0.6$\arcsec diameter for the dropouts selected in the field of RXC J2248. We present only the filters from f625w to f160w since all the dropouts are not detected in the filters bluer than f625. Where the objects are not detected, we give the $5\sigma$ detection limit, determined locally. The last two columns provide the $1\sigma$ upper limits for the IRAC photometry extracted within 1.2\arcsec aperture for the candidates ID2,ID3 and rD.}
\begin{threeparttable}
\begin{tabular}{|l|l|c|c|c|c|c|c|c|c|c|c|}

\hline
\hline
ID&   f625&   f775&               f814 &                  f850  &              f105 &        f110    &    f125   &      f140  &    f160  & 36$\upmu$m& 45$\upmu$m\\
\hline 
ID1&$>25.9$&$> 25.7 $&               $ 25.8\pm0.1 $   & $25.1\pm0.2$  &  $  24.8\pm0.1$  &    $24.8\pm0.1 $ & $   24.8\pm0.1$    & $ 24.9\pm0.1 $ &   $ 24.9\pm 0.1$ &   $  --$           &   $  --$          \\
ID2&$>25.4$&$> 25.4 $&               $ 26.4\pm0.2 $   & $25.0\pm0.1$  &  $  25.2\pm0.1$  &    $25.2\pm0.1 $ & $   25.3\pm0.1$   &  $ 25.4\pm0.1 $ &   $ 25.5\pm 0.1$ &   $ >24.3$ &   $ >24.5$\\
ID3&$>26.4$&$> 26.1 $&               $ 26.1\pm0.1 $   & $25.0\pm0.2$  &  $  24.9\pm0.1$   &   $25.0\pm0.1 $ & $   25.0\pm0.1$   &  $ 25.2\pm0.1 $ &   $ 25.3\pm 0.1$ &   $ >25.6$ &   $ >26.1$\\
ID4&$>26.5$&$> 26.5 $&               $ >27.0$\tnote{a} & $26.0\pm0.3$  &  $  26.0\pm0.1$  &    $26.1\pm0.1 $ & $   25.9\pm0.1$   &  $ 26.1\pm0.1 $ &   $ 26.4\pm 0.2$ &   $ --$            &   $ --$            \\
rD&$>26.4$\tnote{b}&$ 25.9\pm0.1$& $ 25.7\pm0.1 $   & $25.4\pm0.2$  &  $  25.6\pm0.1$  &    $25.6\pm0.1 $ & $   25.7\pm0.1$   &  $ 25.6\pm0.1 $ &   $ 25.8\pm 0.1$ &   $ >27.0$ &   $ >27.3$ \\  
\hline     
\end{tabular}
     
     \begin{tablenotes}
       \item[a] In the filter f814w, ID4 is fainter than the $5\sigma$ detection limit, but is brighter than the $3\sigma$ limit.
       \item[b] In the filter f625w, rD  is fainter than the $5\sigma$ detection limit, but is brighter than the $3\sigma$ limit.
     \end{tablenotes}
\end{threeparttable}
     \label{tab:mags}
\end{table*}

\begin{table*}
\caption{Summary of the photometric redshift computed with \texttt{LePhare} for the 5 dropouts selected in the field of RXC J2248. Col.1 ID; Col.2-3 coordinates; Col. 4-5 first \zphot best solution with the respective $68\%$ confidence level intervals and reduced $\chi^2$ for galaxy fitting; Col. 6-7 QSO \zphot best solution and reduced $\chi^2$ using pure AGN templates; Col. 8 $\chi^2$ for the best  stellar fit; Col. 12 is the first detection band.} 
\begin{tabular}{|c|c|c|c|c|c|c|c|c|c|c|}
\hline
\hline
ID&RA(J2000)&DEC(J2000)&z$\rm_{GAL} [68\% C.L.]$&$\rm\chi^{2}/dof$&z$\rm_{QSO}$&$\rm \chi^{2}_{QSO}$&$\rm \chi^{2}_{Star}$&Det Band\\
\hline
ID1 &  342.18906 &-44.53003 & 5.70  $[5.54-6.01]$&0.9   &6.00&6.5  &10.1 &f814w\\
ID2 &  342.18106 &-44.53461 & 5.88  $[5.72-6.01]$&2.5   &5.88&14.3 &14.2 &f814w\\
ID3 &  342.19089 &-44.53746 & 6.04  $[5.87-6.12]$&1.7   &6.00&20.1 &19.4 &f814w\\
ID4 &  342.17130 &-44.51981 & 5.97  $[5.62-6.32]$&2.0   &6.20&4.3  &6.4 &f814w($<3\sigma$)\\
rD &  342.17145 &-44.54686 & 5.11   $[5.00-5.29]$&2.2   &5.00&9.2 &18.6 &f775w\\
\hline                      
\end{tabular}
     
\label{tab:dropouts}
\end{table*}

\section{Quadruply lensed dropout: photometric evidence}
\label{sec:photometric_evid}
The four \hiz lensed candidates ID1-4 are $\rm i_{f775}$ dropouts:
they are not detected in the filters bluer than the f814w, in which ID1-3 are clearly first detected with detection $>5\sigma$, while ID4 is detected with detection higher than $3\sigma$. Moreover they are detected in all the WFC3IR  bands (see Fig.~\ref{fig:dropouts}). 
The image ID1 is the brightest among the ID1 to ID4, but it is also the one with a highly contaminated photometry since it is close to two very bright 
early-type galaxies (the distance from the centre of the closer galaxy is $\sim1.7\arcsec$). ID4 is the faintest image  and  the furthest from the cluster centre (thus we expect it to be the least magnified), moreover it is close ($\sim1.5''$) to a star.
ID2 and ID3 are the images with the best photometry, since both are in isolated regions: we will refer to these images when using the photometry to estimate physical properties of this system.  
ID2 and ID3 are also identified in both Spitzer/IRAC channels, but the S/N is too low to allow for extraction of robust photometry. Thus we extract photometry in $1.2\arcsec$ aperture and use it as upper limit in our following photometric analysis. About the other two candidates, ID1 is completely embedded in the light of the two nearby galaxies, while ID4, the faintest of our system, is not detected in any of the two channels.

\subsection[]{Photometric redshift}
 \begin{figure*}
     \centering
     \includegraphics[width=18cm]{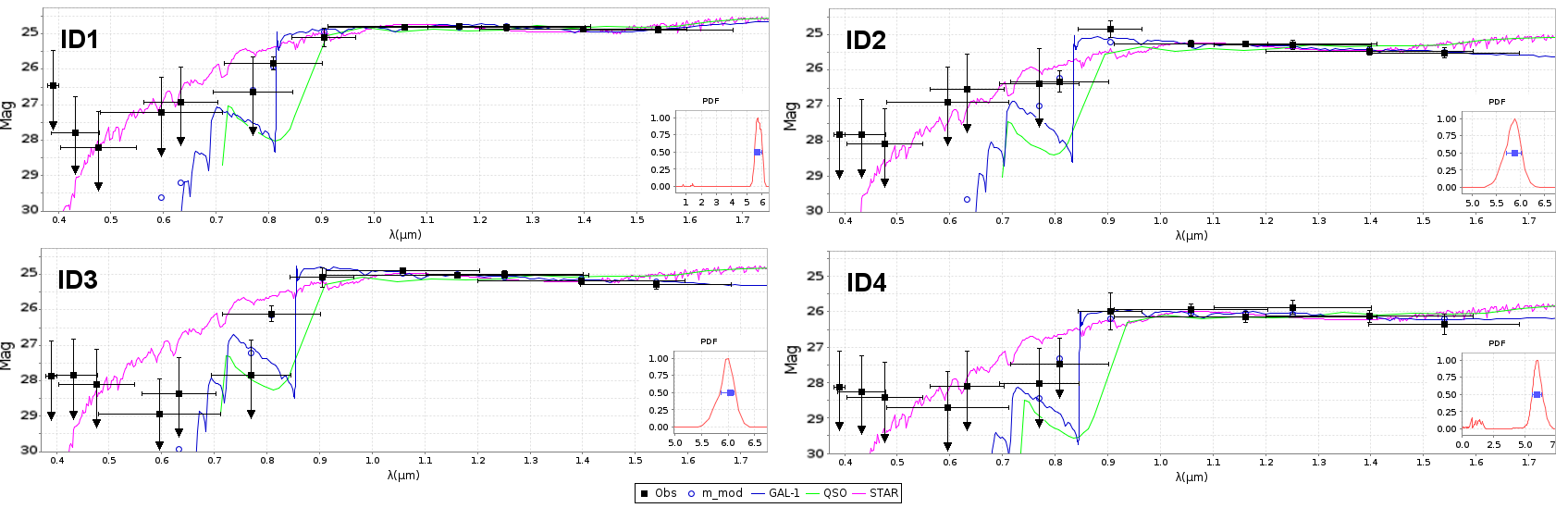}
     \caption{\small SED Fit for the best results from \texttt{LePhare}. The black squares are the observed magnitudes; the open blue circles are the predicted magnitudes; arrows represent upper limits in the detection; the blue, green and magenta lines are the Galaxy, Quasar and Stellar templates respectively.
     The Probability Distribution Function (PDF) of the redshift for the Galaxy template  is given in the lower right corner of each SED-fit.}
   \label{fig:sed_fit}
     \end{figure*}

As described in Sec.~\ref{sec:photo_z}, we compute the photometric redshifts with \texttt{LePhare} using the COSMOS galaxy templates library and photometry within 9 pixels aperture: 
for the ID1-4 dropouts, we obtain only \hiz solution at $z\sim5.9$ with reduced $\rm\chi^2/${\scriptsize DOF} between 0.9 and 2.5, see Tab.~\ref{tab:dropouts}. 
The four \hiz candidates are best fitted with starbust (SB) galaxy templates, the SB10 (for ID1) and SB11 (for ID2,ID3 and ID4) templates from the COSMOS library. ID1 and ID4 best fits are obtained applying the $\rm SB_{calzetti\_bump2}$ extinction law with colour excess E(B-V) of 0.2 for ID1 and 0.1 for ID4. This apparent reddening can be related to the light contamination by the two close early type galaxies for ID1, and to the nearby red galactic star for ID4. The other two objects of our system, ID2 and ID3, require no extinction in the best fit.\\ 
 For the QSO templates, we use the SWIRE QSO library including the two Seyfert templates, see Sec.~\ref{sec:photo_z}.
The best QSO fits are still for a $z\sim6$ source: 
in particular when we include the two Seyfert galaxy SED in the QSO templates, we get a best fit for $ z\rm_{QSO}\sim6$ with $\chi^2/${\scriptsize DOF} lower than the galaxy best fit, while if we only include the pure AGN templates, we still get best $z\rm_{QSO}\sim6$, but with $\chi^2/${\scriptsize DOF} higher than the galaxy $z\rm_{best}$ (see Tab.~\ref{tab:phot}).\\    
As stellar templates we use the Pickles library \cite{Pickles1998}: the stellar SED fits give the worst results, as shown in  Tab.~\ref{tab:dropouts}. 
This, in addition to the extended nature of the candidates, support the exclusion of red dwarf galactic star contaminants.
See Tab.~\ref{tab:dropouts} for a complete summary of the $z_{\mathrm{ph}}$ estimations for the ID1-4 candidates.  \\
As we described, we get only \hiz solutions for our candidates when we run the SED fit on the $z$ range $[0,12]$. Thus to investigate what the values for a low redshift scenario would be, we force the $z_{\mathrm{ph}}$ to be lower than 4 when we run \texttt{LePhare}. With this constraint we get $z_{\mathrm{ph}}$ within 0.8 and 1.4 for the four dropouts, but with  $\chi^2/${\scriptsize DOF} $>10$ due to the bad fit in the NIR filters. 
Actually if we totally exclude the WFC3IR  photometry, using only fluxes up to the f850lp filter,  we get good fits for $z\rm_{ph}\sim1.14$ with reduced $\chi^2/${\scriptsize DOF} within 0.2 and 2.6: in this case the predicted NIR fluxes are $\sim1\,\mathrm{mag}$ brighter than the observed ones.
In Fig.~\ref{fig:sed_fit} we present the best SED fit results from \texttt{LePhare}. The \hiz photometric solutions we get for these sources are in agreement with the  photometric redshift provided in the CLASH  public catalogue\footnote{http://archive.stsci.edu/prepds/clash/} for RXC J2248, except for the ID2 candidate, for which a $z\rm_{ph}\sim0.09$ is obtained. We check the photometry provided in the public catalog for this candidate, and we find that there are fake detections in some of the ACS filters
(f606w, f625w and f775w), due to photometric contamination in these central cluster regions, which leads to the low-z
solution for the \zphot for this candidate. This photometry is also extracted in an aperture of 89 pixels, slightly larger then the aperture we use (64 pixels). In addiction, this candidate is also very close to the  ACS chip gap in the \texttt{Mosaicdrizzle} images (see Fig.~\ref{fig:dropouts}). In our detailed photometric analysis of the ID1-4 candidates, we find that in those filters (and in the bluer ones) ID2 has detection lower than the local
$3\sigma$ detection limit: this, combined with the photometry we
extract in the NIR filters, places the ID2 candidate at $z\sim6$
in our photometric analysis.

\subsection[]{Colour-colour diagrams}
 
\begin{figure*}
     \centering
      \includegraphics[width=17cm]{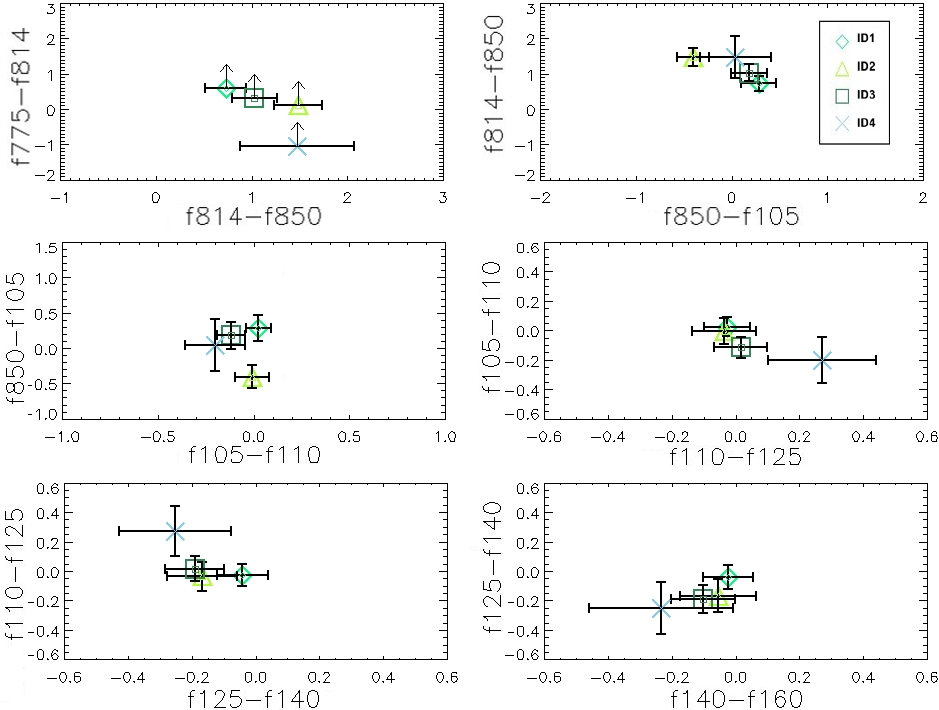}
     \caption{\small HST NIR colour-colour plot of the four \hiz lensed candidates using  photometry in 0.6'' aperture.
     In the first panel we show the colours over the dropout filter (the arrows represent lower limit). The colours are consistent within $4\times$ the formal \texttt{SExtractor} errors, which correspond to our estimated true photometric errors in the aperture (see Sec.~\ref{sec:Dataset}).}
   \label{fig:drop_col_col}
     \end{figure*}

To assure the robustness of dropout \hiz candidates and to rule out low-$z$ contamination, namely by early-type galaxies at $z\sim1.3$, we analyse the NIR colours of our \hiz candidates in the colour-colour diagram.
In the following colour analysis of the z$\sim6$ dropout, we refer to the photometry in the dropout filter, f775w, and in the NIR filters, f105w, f125w, f160w. We do not consider any impact of the Ly$-\alpha$ emission line contribution, since  the  z$\sim6$ photometric redshift predicted for ID1-4 places the Ly-$\alpha$ emission in the f814w and f850lp filters.
In Fig.~\ref{fig:col_col_diagr} we show the optical and NIR colours diagram for the sources extracted in the FOV of RXC J2248, the \hiz dropouts that we select in this field and the colour$-$redshift tracks of star-forming galaxies (with $\beta=-1,-2,-3$) with which we compare the colours of our dropouts. In addiction, we also plot the colour$-$redshift track for the SB template from the COSMOS library which best fit our candidates (SB11). 
Our four candidates ID1-4 all lie in the region defined with the colour criteria of Eq.~\ref{eq:color_criteria}: their observed NIR colours are consistent with the colours predicted for \hiz star forming galaxies with UV slope $\beta$ between -2 and -3. 
We measure the UV-slope from the observed NIR photometry (see \citealt{Bouwens2012},\citealt{Bouwens2013}): 
\begin{equation}
\rm \beta=3.09((Y_{105}+J_{125})/2-H_{160})-2
\label{eq:beta}
\end{equation}
and we find that our candidates have a mean $\beta=-2.89\pm0.38$, which is slightly bluer than the mean UV slope $\beta=−2.24\pm0.11\pm0.08$  measured for $z\sim6$. 
\cite{Bouwens2013} shows that measurements of $\beta$  are very sensitive to accurate photometry, and that bluer estimations can be due to errors in PSF estimations. To investigate such effect, we PSF match our nir images to the f160 image using the \texttt{psfmatch} routine in IRAF\footnote{http://iraf.noao.edu/}. We get $\beta=-2.88\pm0.49$, thus the PSF matching has no effect on our $\beta$ estimation.\\    
Since lensing is  achromatic, multiple images of a source are expected to have the same colours, unless the source itself has intrinsic spatial colour variation, or the extinction along the line of sight is different for the different multiple images. Thus, we compare all the NIR colours of the 4 dropouts with each other, to support their lensed nature on the basis of photometry. As shown in Fig.~\ref{fig:drop_col_col}, the colours of the 4 dropouts are all consistent within 4 times the \texttt{SExtractor} formal errors,  with the only exception of the f105-f110 colour for ID4, which is slightly greater than the other dropouts.
     
\section[]{Quadruply lensed dropout: lensing evidence}
\label{sec:lensing_evid}
Given the positions of the ID1-4 candidates in the field of the cluster (see Fig.~\ref{fig:rgb}), we investigate the possible multiply lensed nature of the system. Therefore we now derive a strong lensing model of RXC J2248 to verify quantitatively whether they are lensed images of the same background source.\\ 
We model the inner mass profile of the cluster  with the strong lensing modelling software \texttt{GLEE} developed by S. H. Suyu
and A. Halkola \citep{Suyu2010,Suyu2012}. Given information as 
positions and redshifts of multiple images and using parametric
models to describe the mass profiles of the lenses, the code yields
 the best fitting model which reproduces the observed images (through a
simulated annealing minimisation in the source and/or image planes),
as well as a Monte Carlo Markov Chain (MCMC) sampling to find the most
probable parameters and uncertainties for the model. 
 
\subsection{Mass components}
 We describe the smooth dark halo (DH) mass component of the cluster with a Pseudo Isothermal Elliptical Mass Distribution (PIEMD) profile \citep{Kassiola1993}, with projected surface density given by

 \begin{equation}
 \Sigma(R)= \frac{\sigma^2}{2 {\rm G}}(R_{\rm c}^2+R^2)^{-0.5}
 \end{equation}
where $\sigma$ is the velocity dispersion of the DH and $R\rm_c$ is the core radius.  
The galaxy component is modelled with truncated singular isothermal elliptical profiles (BBS) \citep{Brainerd1996}, parametrized by the central velocity dispersion $\sigma$ and the truncation radius $r_t$. 

The projected surface mass density  for this profile is:
 \begin{equation}
 \Sigma(R)= \frac{\sigma^2}{2 {\rm G} R} \left[ 1 -
   \left(1+\frac{r_{\rm t}^2}{R^2}\right)^{-0.5} \right] \quad .
 \label{theory:BBS-2D-density}
 \end{equation}

 We select 80 cluster members  to include in our strong lensing model, counting 16 spectroscopically confirmed cluster members from \cite{gomez2012} and 64  cluster members selected through photometric redshift and magnitude-colour cuts  (see Sec.~\ref{sec:photo_z}). \\
As in \cite{Halkola2006} and \cite{Eichner2013}, to reduce the number of free parameters, we adopt luminosity scaling relations for the velocity dispersion $\sigma$ (based on the Faber-Jackson relation) and the truncation radius $r_t$ (as measured in \cite{Eichner2013}) of the galaxies: 
\[\sigma=\sigma^*\left(\frac{L}{L^*}\right)^{\delta} \quad\quad r_{\rm t}=r_{\rm t}^*\left(\frac{L}{L^*}\right)^\alpha
\]
where the amplitudes $\sigma^*$ and $r_{\rm t}^*$ are the velocity
dispersion and halo size for a reference galaxy halo with luminosity
$L^*$, while the exponents of the luminosity relations are $\delta=0.25$ and $\alpha=0.5$.  Assuming that $r_{\rm t}$ scales
with $\sigma$, $r_{\rm t}=r_{\rm t}(\sigma)$ \citep{Halkola2006,Hoekstra2004}, the free parameters used to tune the galaxy mass contribution are reduced to the reference galaxy velocity dispersion $\sigma^*$ and truncation radius $r_{\rm t}^*$. In this work we use as reference galaxy the second brightest galaxy of cluster. 
The amplitudes of the luminosity relations are measured for both cluster galaxies in the cluster core \citep{Eichner2013} and field galaxies \citep{Brimioulle2013}. However, peculiar galaxies, as the cluster BCGs,  have large scatter from these luminosity relations (see \citealt{Postman2012b}, \citealt{Kormendy2013}) and likely a different size relation; therefore we need to optimise them independently to robustly account for their contribution to the total mass profile.\\
We use the observed fluxes in the f814w band as a tracer for
the luminosity $L$ of the cluster members. For the ellipticities and orientations we use the values estimated with \texttt{SExtractor} in the same band (assuming that the dark matter halo ellipticities of the cluster galaxies  are equal to their surface  brightness ellipticities).
\subsection{Multiple images}

Since we lack spectroscopic confirmation of multiple image systems, we use the wide CLASH photometric dataset to select multiple images based on similarity of surface brightness,
morphology and photometric redshift: we identify 13 candidate multiple lensed systems, for a total of 37 multiple images, without counting the high redshift lensed candidate system (see Fig.~\ref{fig:rgb}). The sources redshifts $\rm z_s$ are free parameters in our models: for each system, we use the mean photometric redshift of the multiple images as starting value for $\rm z_s$, and we optimise it within an interval of dz$\sim\pm0.5$ using a gaussian prior (for the system 12, since we get different \zphot for the multiple images, we use a larger interval of dz$=\pm1$).
\cite{Host2012} and \citet{D'Aloisio} show that, on cluster scales, multiple image positions can be reproduced with accuracy not better than 1-2 arcseconds due to density fluctuations along the line of sight. Starting with realistic uncertainties on the multiple images positions is important in the modelling procedure, because  underestimating such errors will lead to underestimated uncertainties on the mass model free parameters. Thus we adopt a positional uncertainty of $1\arcsec$ on the  multiple images positions, even though from the HST images we can estimate such positions with an accuracy of $0.065\arcsec$.\\ In Tab.~\ref{tab:multiple images} we list all the multiple lensed systems  with the respective positions and \zphot.
\begin{table}
\caption{Summary of the multiple lensed systems used to constrain the strong lensing model of RXC J2248.
  The columns are: Col.1 is the ID; Col.2-3 Ra and Dec; Col.4  is the photometric redshift from \texttt{LePhare}, Col.5 is the redshift predicted from the lensing model. }
\footnotesize
\begin{tabular}{l|l|l|l|l|}
\hline
\hline
 Id & Ra & Dec & $\rm z_{ph} [68\% C.L.]$& $\rm z_{lens} [68\%C.L.]$ \\
 \hline 
 1a       & 342.19585  &  -44.52889    &  1.22   [1.12-1.25]    &   1.22  [1.17-1.27] \\ 
 1b       & 342.19450  &  -44.52702    &  1.09   [1.07-1.11]    &  1.22  [1.17-1.27]\\
 1c       & 342.18646   & -44.52119    &  1.28   [1.27-1.29]    &  1.22  [1.17-1.27]\\
 \hline
 2a       & 342.19560   & -44.52843    &  1.11   [1.12-1.12]    &   1.23  [1.18-1.28] \\
 2b       & 342.19479  &  -44.52732    &        --             &1.23  [1.18-1.28]\\      
 2c       & 342.18630    & -44.52107   &  1.23   [1.22-1.25]    &   1.23  [1.18-1.28]\\  
 \hline
 3a       & 342.19257  & -44.53073    &  1.29   [1.27-1.31]    &   1.27  [1.22-1.31]\\
 3b       & 342.19247  & -44.53046     &  1.24   [1.22-1.25]    &    1.27  [1.22-1.31]\\
 3c       & 342.17974  & -44.52157     &  1.13   [1.11-1.13]    &    1.27  [1.22-1.31]\\
 \hline
 4a       &   342.19315  &-44.53653    &  1.29   [1.27-1.30]    &   1.43  [1.36-1.50]\\
 4b       &   342.18781  &-44.52732    &  1.49   [1.43-1.54]    &   1.43  [1.36-1.50]\\
 4c       &   342.17919  &-44.52353    &  1.40   [1.39-1.41]    &   1.43  [1.36-1.50]\\
 \hline
 5a        &  342.19294  &-44.53659    &   1.37  [1.35-1.38]   &    1.43  [1.36-1.51] \\
 5b        &  342.18774  &-44.52747    &   1.38  [1.36-1.39]   &     1.43  [1.36-1.51] \\
 5c        &  342.17889  &-44.52361    &   1.45  [1.24-1.48]   &     1.43  [1.36-1.51] \\
 \hline
 6a       &  342.18843    & -44.54002  &  1.38   [1.36-1.40]  &     1.44  [1.38-1.51]\\
 6b       &  342.17580    & -44.53253  &  1.69   [1.68-1.68]  &     1.44  [1.38-1.51]\\
 6c       &  342.17417   &  -44.52837  &  1.53   [1.51-1.56]  &     1.44  [1.38-1.51]\\
 \hline
 7a       &  342.18005   &  -44.53842  &  0.92   [0.90-0.94]   &   1.04  [0.99-1.08]\\
 7b       &  342.17553     & -44.53589 &    --                 &    1.04  [0.99-1.08]\\
 7c       &  342.17193   &  -44.53024  &  1.05   [0.04-1.07]   &  1.04  [0.99-1.08]\\
 \hline
 8a      &  342.18187   &  -44.54049  &   2.18   [2.16-2.20]  &    1.96  [1.87-2.05]\\
 8b      &  342.17427   &  -44.53710  &   2.12   [2.10-2.14]  &   1.96  [1.87-2.05]\\
 8c      &  342.16940   &  -44.52722  &   2.12   [2.08-2.15]  &   1.96  [1.87-2.05]\\
 \hline
 9a      &  342.18032   &  -44.54083  &   2.92   [2.82-2.99]  &    2.91   [2.71-3.14]\\
 9b      &  342.17476   &  -44.53858  &   3.02   [2.91-3.08]  &   2.91   [2.71-3.14]\\
 9c      &  342.16780   &  -44.52628  &   2.70   [2.55-2.79]  &   2.91   [2.71-3.14]\\
 \hline
 10a      & 342.18078    & -44.54089  &  2.94   [2.96-3.05]   &   2.95  [2.75   3.20]\\
 10b      & 342.17462    & -44.53842  &  3.12   [2.99-3.26]   &  2.95  [2.75   3.20]\\
 10c      & 342.16792     &-44.52621  &                        --&   2.95  [2.75   3.20] \\
 \hline
 11a     &  342.17504   &  -44.54102  &   3.12  [3.04-3.19]   &   2.95  [2.86-3.04]\\
 11b     &  342.17315    &  -44.53999 &   2.91  [2.83-2.99]   &   2.95  [2.86-3.04]\\
 11c     &  342.16557     &-44.52954  &   3.04  [2.64-3.18]   &   2.95  [2.86-3.04]\\
 \hline      
 12a       & 342.19555 & -44.53213    &   1.91  [1.75-2.54]    &  2.27  [1.99-2.66]\\
 12b       & 342.19392 & -44.52870    &   1.75  [1.46-1.90]   &  2.27  [1.99-2.66]\\
 12c        & 342.18278 & -44.52152   &   3.06  [2.96-3.15]   &  2.27  [1.99-2.66]\\
 \hline   
 13a    & 342.19369 & -44.53012       &   1.59  [1.13-2.24]  & 1.371  [1.25-1.65]\\
 13b    & 342.19333 & -44.52941       &   1.57  [1.12-3.21]  & 1.371  [1.25-1.65]\\

\hline
 \hline 
  \end{tabular}\\
\label{tab:multiple images}
\end{table}

\subsection{Results}
\begin{figure*}
      \centering
      \includegraphics[width=18cm]{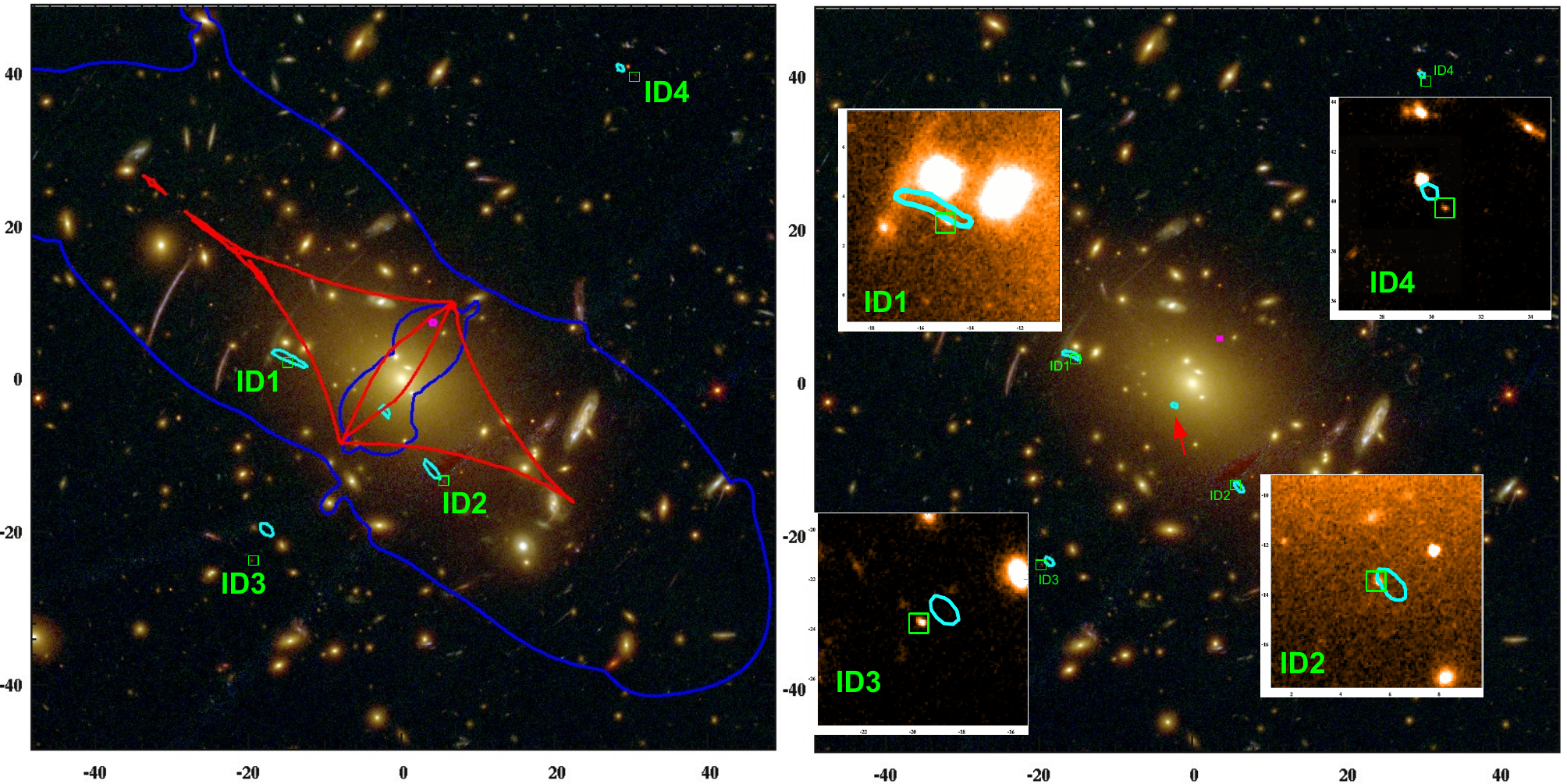}%}
      \caption{\small HST colour composite images of the inner region
        of RXC J2248. \textbf{Left Panel}: critical lines (in blue) and
        caustics (in red), for a source at z$=6$, for the best strong lensing model that we obtain using as constraints the 13 multiple
        image systems given in Tab.~\ref{tab:multiple images} (excluding the ID1-4 \hiz system). The light green squares show
        the ID1-4 positions, while in light blue we show the images that this
        model predicts for a source at z$=6$ with a radius of $0.25\arcsec$. \textbf{Right Panel}:
        Images prediction for the system ID1-4 from the final best
        model, in which we also use this \hiz system as constraint in the modelling: the green squares are the observed positions and the light blue contours are the predicted images for a source with size of 0.25\arcsec. From this model we get z$_s=6.1_{-0.4}^{+0.3}$. We show the zoom on the
        predicted images ($8.0\arcsec\times8.0\arcsec$) in the
        HST/f110w image. The red arrow indicate the position of the
        5th central image predicted for this system.}
    \label{fig:sl_mod1}
      \end{figure*}

First we optimise the mass model using the 13 multiple image systems in Tab.~\ref{tab:multiple images} as constraints, getting the best fitting model which reproduces the positions of the observed multiple lensed images
 with $\chi_{\rm tot}^2=16.2$ (having 21 degrees of freedom) in the image plane. Then we run the MCMC to sample the most likely parameters for the model.\\ We use this mass model of RXC J2248 to analyse the multiple imaging for a source at $z\sim6$ 
and we find that the model predicts a 4 multiple images configuration very close to the one that we observe 
 (see Fig.~\ref{fig:sl_mod1}).
We also test the prediction assuming that the source of this system is at low-$z$. From \texttt{LePhare}, when we force the photometric redshift of the ID1-4 images to be at low-$z$,  we get the $z_{\mathrm{ph}}$  to be within 0.8 and 1.4 when (see Section~\ref{sec:photometric_evid}). For redshift of the source within this range, the model predicts lensed images which have no correspondence with sources observed in the field of the cluster, thus supporting that the lensed system is at high redshift.\\ In addition, we verify that independent mass modelling with the full light-traces-mass method of \citealt{Zitrin2009} supports the identification of multiple images outlined here, as well as the $z\sim6$ solution for the quadruply lensed source.\\%

\begin{figure}
      \centering
      \includegraphics[width=8.5cm]{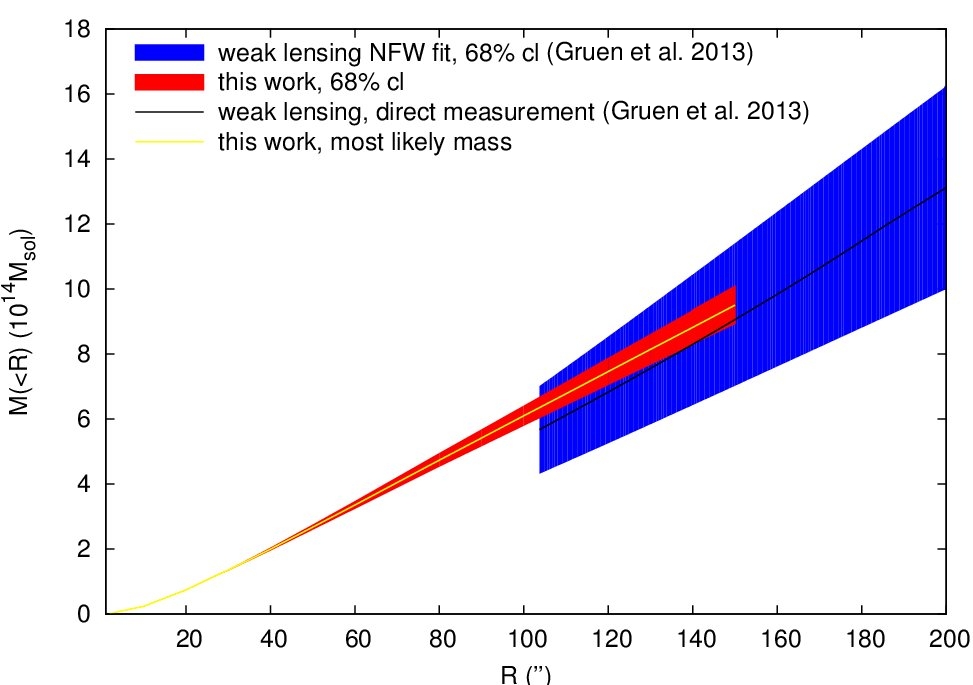}%}
      \caption{\small Cumulative projected mass profile of RXC J2248. We compare results from the strong lensing (this work) and weak lensing \citet{Gruen2013} analysis. In yellow we plot the most likely mass from the strong lensing analysis with $1\sigma$ uncertainties (red area) obtained  from the MCMC sample. In black we plot the weak lensing measurements from \citet{Gruen2013} with the uncertainties from the best NFW fit model (blue area).  The predictions from both lensing analyses are in good agreement in the radial region where they overlap. }
    \label{fig:mass_sl_wl}
      \end{figure}

Once we tested that our model predicts the observed \hiz multiple
lensed system, we perform a new strong lensing model of the cluster,
using this time also the \hiz quadruply lensed system ID1-4 in the modelling. For the redshift of this system we use a starting value of $z=6$ and allow it to vary within the range $z=[5.5,6.5]$ throughout the modelling process.
After the minimisation and the MCMC runs, we get a final model
of the cluster with $\chi^2=22$ with 27 degrees of freedom.  
For the cluster--scale halo, we get an offset of $\delta x=-0.6\arcsec_{-0.7{\tiny\arcsec}}^{+0.6{\tiny\arcsec}}$, $\delta y=0.7\arcsec_{-0.5{\tiny\arcsec}}^{+0.5{\tiny\arcsec}}$ with respect to the BCG,
 axis ratio of $b/a=0.5_{-0.03}^{+0.03}$ and position angle
P.A.=$-37.8^{\circ+0.7^\circ}_{\,\,\,-0.9^\circ}$. The cluster mass distribution has a core radius of $\sim16.9_{-2.2}^{+2.5}\arcsec$, and it has an Einstein radius $\theta_E$=$29.9\arcsec_{-1.9{\tiny\arcsec}}^{+1.7{\tiny\arcsec}}$ for a source at $z_{\rm s}=2$, which gives a central velocity dispersion of
$\rm\sigma= 1018_{-45}^{+40}\,km/s$ for a singular isothermal sphere.
The BCG is well centred on the DH, its axis ratio is $b/a=0.84\pm0.02$, the major axis has an offset of $\sim 20^{\circ}$ with respect to the DH orientation, and has a velocity dispersion of $\rm \sigma=286_{-147}^{+173}\,km/s$. 
The total mass of the cluster  is $M_{\rm tot}=1.24\pm0.01\times10^{14}M_\odot$ within the Einstein radius for a source at $z_{\rm s}=2$. To compare our mass prediction with the weak lensing results from \citet{Gruen2013}, we calculate the most likely mass and its $1\sigma$ uncertainties by taking 200 random MCMC sample points. 
We derive the surface mass density and measure the projected mass up to $2.5\arcmin$ for each model and measure their projected mass within different apertures. The prediction from both lensing analyses are in good agreement in the radial region where they overlap, (see Fig.~\ref{fig:mass_sl_wl}). 
Our best model of the cluster predicts the 4 multiple lensed \hiz system with an accuracy of $0.8\arcsec$ on the observed positions and a source redshift of $6.1_{-0.4}^{+0.3}$. In Tab.~\ref{tab:multiple images} we list the $z_s$ predicted by our final best model also for all the other systems of multiple images. 
 The magnifications induced by the cluster for the ID1-4 system
are $\mu_{\rm ID1}=8.3\pm3.9$, $\mu_{\rm ID2}=5.1\pm1.0$, $\mu_{\rm ID3}=5.3\pm0.6$ and $\mu_{\rm ID4}=2.2\pm0.2$. The magnification of the brighter image, ID1, is not well constrained due to its position very close to the critical lines.  In proximity of the critical lines, the magnification gradient rapidly vary, leading to high uncertainties in the magnification in correspondence of small uncertainties in the position.  
We estimate the relative magnification of ID2 and ID3 to be $\mu_{23}=0.95\pm0.10$,  which is in agreement with the observed fluxes ratio within 4 times the \texttt{SExtractor} photometric errors. \\
Finally, comparing the previous model, and the final one with the \hiz system, we found that including or omitting the high redshift candidate system does not significantly alter the results for the cluster mass model.

\section{The central image}
\label{sec:central_image}
\begin{figure*}
     \centering
      \includegraphics[width=18cm]{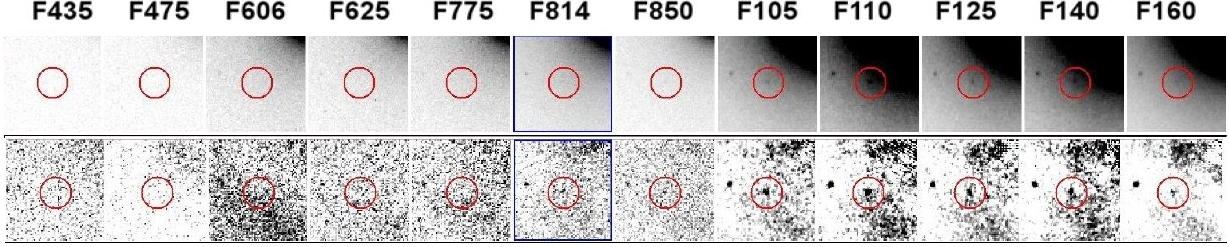}
     \caption{\small Postage stamps of the central image that we discover thanks to the strong lensing prediction and the subtraction of the cluster BCG.
     \textbf{Upper panel}: HST/ACS and HST/WFC3IR postage of $4\arcsec\times4\arcsec$  centred on the position predicted from lensing for the 5th image of the \hiz system, where a source is actually observed in the NIR images.
     \textbf{Lower panel}: same as the upper panel, but for the images after the BCG subtraction: in this case there is a clear detection in the NIR filters, but also in the f814w and f850lp filters there are residuals left after the BCG subtraction.}
   \label{fig:central_image}
     \end{figure*}

As we show in Fig.~\ref{fig:sl_mod1}, our final best model predicts a central fifth image for our \hiz system, as we also expect from lensing theory. This central image is $\sim 3.5\arcsec$ away from the BCG, thus it is completely embedded in the BCG light. However,
 we find that, in the NIR images of the CLASH dataset, a source is detected at the position predicted for the central image (see Fig.~\ref{fig:central_image}). We call this source ID0.
To investigate this source we subtract the BCG light from our images. Within CLASH, we use the isophote fitting routine, SNUC\footnote{see http://astronomy.nmsu.edu/holtz/xvista/index.html and \citealt{Lauer1986}}, which is part of the XVISTA image processing system , to derive two-dimensional models of all early-type BCGs in the CLASH clusters. SNUC is capable of simultaneously obtaining the best non-linear
least-squares fits to the two-dimensional surface brightness distributions in multiple, overlapping galaxies (see \citealt{Lauer1986}). We perform these fits independently in the 12 HST passbands acquired with the ACS/WFC and WFC3/IR cameras (the BCGs are typically not dominant in the four WFC3/UVIS bands used in CLASH). Up to ten of the brightest galaxies, including the BCG, are fitted. Other objects in this region are masked prior to the fit.\\ In the specific case of RXC J2248, we only fit the BCG and its three closest satellite galaxies. Residuals were typically less than 0.001 mag, indicating that all 4 of the fitted galaxies are consistent with the concentric elliptical isophote assumption used in this procedure. SNUC, however, does allow the ellipticity and position angle of the isophotes to vary with the semi-major axis.\\ In Fig.~\ref{fig:central_image} we show the cutout of the central image ID0 of our lensed system for the HST/ACS and HST/WFC3IR images after the subtraction of the BCG and the satellite galaxies.
After removing the BCG, ID0 is clearly detected in all the NIR HST filters, while its  detection in the f814w and f850lp filters is difficult to claim, given the high noise fluctuations in this two bands and their low S/N.
Despite the fact that the photometry of this object is highly contaminated by the BCG light, we try to estimate the \zphot using the photometry extracted from the images after the BCG subtraction. The result from \texttt{LePhare} gives \zphot$=5.88$ as the best solution for the galaxy SED fit.
Due to the noisy photometry, the PDF($z$) is quite broad, but given the good lensing prediction for this image, and also the best \zphot value, we are confident that this source is actually the central image of our system.       
For this image, our best strong lensing model predicts a magnification of $2.12\pm0.68$. From the lensing model, the magnifications of ID2 and ID3  relative to ID0 are $\mu_{20}=2.4\pm1.2$ and $\mu_{30}2.5\pm0.9$ respectively. These magnifications are in agreement within the uncertainties with the mean ratios of the observed fluxes in the NIR filters (from f105w to f160w), which are $\rm f_{20}=1.1\pm0.3$ and $\rm f_{30}=1.3\pm0.3$.

\section{Physical properties}
\label{sec:Properties}
In this section we estimate some physical properties of the \hiz galaxy quintuply lensed in the field of RXC J2248.
As noted in the previous sections, the images ID2 and ID3 are the ones with the  best photometry, which do not suffer any contamination by nearby sources, so we focus on these images to estimate physical properties based on photometry.
In Sec.~\ref{sec:lensing_evid}, we estimate the magnification for ID2 and ID3 to be  $\mu_{\rm ID2}=5.1\pm1.0$ and $\mu_{\rm ID3}=5.3\pm0.6$. From the best \texttt{LePhare} galaxy SED fit, we extract the UV absolute magnitude $\rm M^{AB}_{1600}$ for these images, and after correcting for the magnification, we get delensed absolute UV magnitudes of $-19.6\pm0.2$ and $-19.8\pm0.1$ respectively for ID2 and ID3. 
 Comparing these values with the luminosity function at redshift z=6 as estimated in \cite{Bouwens2012}, our candidates have $\rm L_{UV}\sim0.5L^*$.\\
 
In order to derive the physical properties of the system ID2\&3 we perform a fit of model SEDs
to the colours of ID2\&3 with the SED-fitting routine \texttt{SEDfit} \citep{2004ApJ...616L.103D}.
It performs a $\chi^2$ fit of model SEDs to the observed photometry, concurrently allowing reddening by dust following the law of \citet{2000ApJ...533..682C}.
As basis for the template set for the fitting we use the single stellar population (SSP) model SEDs of \citet[][hereafter BC03]{2003MNRAS.344.1000B} with a Chabrier IMF \citep{2003PASP..115..763C} and Padova 1994 evolutionary tracks.
From these we create composite stellar populations (CSPs) with the software \texttt{galaxev} \footnote{\url{http://www.cida.ve/\textasciitilde bruzual/bc2003} or \url{http://www.iap.fr/\textasciitilde charlot/bc2003}}.
The star formation history (SFH) of galaxies is commonly described by a so-called '$\tau$ model' that follows:
\begin{equation}
 \mathrm{SFR}\propto\exp(-t/\tau),
\end{equation}
where $SFR$ denotes the star formation rate, and $\tau$ the (positive) $e$-folding timescale.
$t$ is the time that has elapsed since the start of star formation, i.e., the age of the galaxy.
This function describes well the SFHs of local galaxies, but likely does not hold for $z\gtrsim1$ \citep[e.g.,][]{2010MNRAS.407..830M,2013arXiv1303.5059L}.
In fact, \citet{2010MNRAS.407..830M} showed that fitting model SEDs with exponentially decreasing SFRs to star-forming galaxies at $z\sim2$ yields unrealistic young ages because the galaxy spectrum is then dominated by the young stellar population.
Moreover, they showed that $\tau$ models with negative $\tau$ lead to more physical results.\\
Here we create CSPs for $\tau$ models with both positive and negative $\tau$ (and therefore decreasing and increasing SFR), as these should set lower and upper limits to the SFR of the investigated galaxy.
We generate CSPs with metallicities of $Z=0.0001$, $0.0004$, $0.004$, $0.008$, $0.02\ (Z_\odot)$, $0.05$ and $e$-folding timescales of $\tau=\pm0.01$, $\pm0.1$, $\pm0.5$, $\pm1.0$, $\pm2.0$, $\pm3.0$, $\pm4.0$, and $\pm5.0\,\mathrm{Gyr}$ for the $\tau$ model.
The created CSPs were extracted at $24$ different ages evenly distributed in logarithmic space between $0.1\,\mathrm{Myr}$ and $3\,\mathrm{Gyr}$.
Additionally to the CSPs, we extract SSPs with the same metallicities and ages.
We set the redshift of the system ID1-4 to $z=5.9$.
For model ages that are higher than the age of the universe at redshift $5.9$ the SED fitting code will assign probabilities of zero.
The extinction is allowed to take values between $A_\mathrm{V}=0.0$ and $3.0$ with steps of $0.1$.\\
We perform the SED fitting for model sets containing the SSPs and CSPs with increasing and decreasing SFH separately, and afterwards with all models combined.
The results are summarised in Tab.~\ref{tab:SEDfitting}.\\
Using the SSPs models, the best fits yield subsolar metallicities (0.2 and $0.005\,Z_\odot$ for ID2 and ID3 respectively) and very young ages (0.1 and 1.5 Myr for ID2 and ID3 respectively). 
The results when using the CSPs with increasing and decreasing SFR are similar to one another.
The best fits yield the same values for the metallicities (0.2 and $0.005\,Z_\odot$) and comparable small ages ($0.5$ and $1.5\,\mathrm{Myr}$).
In summary, for the three different model sets (SSPs, and CSPs with $\tau\lessgtr0$) the results for metallicities, extinctions, and ages are essentially the same (except for the age of ID2, which is $0.1\,\mathrm{Myr}$ for the CSPs, and $0.5\,\mathrm{Myr}$ for the SSPs).
Actually this is what we would expect given the small ages resulting from the best fits.
Within these short timescales ($t\ll\tau$) the galaxies could not evolve significantly which is why the stellar populations (and therefore SEDs) of the CSPs are very similar to one another and also to the SSPs.
When including all models (CSPs and SSPs) in the SED fitting, we get that ID2 is better fitted by a CSP model with increasing SFR, while an SSP model is preferred for ID3.
In any case, the differences in the stellar populations are, as mentioned above, not very high at these young ages which is why the results for ID2\&3 are in good agreement.\\
All the results are shown in greater detail in App.~\ref{app:sedfit}, Figs.~\ref{fig:incSFH} to \ref{fig:allSFH}, where we plot the best fitting SEDs and the likelihood distributions in parameter space, as well as in Figs.~\ref{fig:PageincSFH} to \ref{fig:PageallSFH}, which display the PDFs of the model ages.
Through interpolation of the age PDF, we calculate that ID2\&3 have ages within the interval of [0.1,330]\,Myr (centred on the PDF) at a $95.45\,\%$ ($2\sigma$) confidence.
The same was done for the masses, which we estimate to be within $\rm [0.3,7.5]\cdot10^{8}M_{\odot}$ at the same level of confidence.
The best fitting masses and the age and mass intervals are also summarised in Table~\ref{tab:SEDfitting}. 
The $\chi^2$ values of all fitting results lie between 10 and 12.7 and differ only marginally when the underlying model set is changed.\\
We furthermore estimate the UV slope $\beta$ (Tab.~\ref{tab:SEDfitting}) of the best fitting SED of each run by a linear fit of $\log(\lambda)$ versus $\log(F_\lambda)$ within $\lambda\in[1276,2490]$\,\Ang\, \citep[see][]{Calzetti1994}, getting results which are in total agreement with $\beta=-2.89\pm0.38$ estimated using the observed NIR colours in Sec.~\ref{sec:photometric_evid}.\\
We also combine the likelihoods from the SED fits of the two candidates, and the results for masses and ages are same as the previous ones. 
Moreover, we repeat the SED fits with SSP and CPS, using this time the combined photometry of the two lensed images ID2\&3, and the SED fits lead to same constraints on masses and ages. \\
\begin{table*}
\caption{Best fitting parameters from SED fitting for CSPs, SSPs, and all models combined.
Positive values for $\tau$ stand for an exponentially decreasing SFR, $\mathrm{SFR}\propto\exp(-t/\tau)$, while negative values denote an increasing SFR.
We interpolate the PDF of model ages (see also Figs.~\ref{fig:PagedecSFH} to \ref{fig:PageallSFH} in App.~\ref{app:sedfit}) and calculate the age interval within which the probability of a fit is $95.45\,\%$ (corresponding to a $2\sigma$ confidence interval).
These $2\sigma$ intervals are given in brackets in the column ``age''.
The same is done for the masses, where the corresponding intervals are in column ``$M_{*}$''.
Note that the age and mass values of the best fitting model do not necessarily lie in the calculated intervals.}
\centering
\footnotesize
\begin{tabular}{c|c|c|c|c|c|c}
\hline
\hline
ID & $\tau$ / Gyr & $Z\ /\ Z_\odot $ & age  [age] / Myr & $M_{*}\ [M_{*}]$ / $10^9\,M_{\odot}$  &  $\beta_{UV}$ & $A_{\mathrm{V}}$\\
\hline
 \multicolumn{7}{c}{\textit{SSPs}}\\
 \hline
 % ID   tau    Z        age     [age]         M        [M]           beta       AV
 ID2    & SSP  &  0.2   &  0.1  [0.1,45] &  0.093 [0.17,0.28] & $-2.90\pm0.02$ & 0.2\\
 ID3    & SSP  &  0.005 &  1.5  [0.1,111]&  0.21 [0.034,0.79]  & $-2.64\pm0.02$ & 0.4\\
 \hline        
 \multicolumn{7}{c}{\textit{$\tau<0$ model CSPs}}\\
 \hline
 % ID   tau    Z         age     [age]         M        [M]           beta      AV
 ID2    &  -1.5 & 0.2   & 0.5  [0.1,274] & 0.093  [0.034,0.25] & $-2.90\pm0.02$  & 0.2\\
 ID3    &  -0.1 & 0.005 &  1.5 [0.1,301] & 0.23   [0.059,0.71] & $-2.62\pm0.02$  & 0.4\\
 \hline
 \multicolumn{7}{c}{\textit{$\tau>0$ model CSPs}}\\
 \hline
 % ID   tau    Z        age     [age]         M        [M]           beta       AV
 ID2    &0.1 &  0.2   &  0.5  [0.1,291]  & 0.093  [0.034,0.27] &  $-2.90\pm0.02$ & 0.2\\
 ID3    &0.1 &  0.005 &   1.5  [0.1,383] & 0.23   [0.059,0.78] &  $-2.62\pm0.02$ & 0.4\\
 \hline   
 \multicolumn{7}{c}{\textit{all combined}}\\
 \hline
 % ID   tau    Z         age     [age]         M        [M]           beta      AV
 ID2    & -1.5 & 0.2   &  0.5 [0.1,270]   & 0.093  [0.033,0.26] & $-2.90\pm0.02$ & 0.2\\
 ID3    &  SSP & 0.005 &   1.5  [0.1,330] & 0.21   [0.058,0.75] & $-2.64\pm0.02$ & 0.4\\
\hline  
\end{tabular}
\label{tab:SEDfitting}
\end{table*}
Finally we investigate whether the inclusion of the additional information from the shallow IRAC data affects our results. We repeat the previous SED fitting but include  the upper limits estimated in the 3.6\,$\upmu$m and 4.5\,$\upmu$m filters in the photometry. In this step we only focus on ID3, since the ID2 IRAC photometry is significantly contaminated by the near cluster members.  
The best fitting SSP model has an age of 1.5\,Myr within the  $2\sigma$ interval of [0.1,30]\,Myr and mass of $\rm2.2\cdot10^8\,M_{\odot}$ within [0.4,5.2]$10^8\,M_{\odot}$. For the increasing and decreasing SFH we get respectively $\tau=-2.5$\,Gyr and $\tau=2.5$\,Gyr, ages of 2.5\,Myr within [0.1,222]\,Myr and 2.5\,Myr within [0.1,212]\,Myr, and masses of  $\rm2.2\cdot10^8\,M_{\odot}$ within [0.6,4.7]$10^8\,M_{\odot}$.  In these cases, the best fitting models require a higher $\tau$ then the previous results, in which we get $\tau=\pm0.1\,Gyr$. However, in the appendix (Figs.~\ref{fig:incSFH} and \ref{fig:decSFH} for increasing and decreasing SFH models) we show that the probability distributions of $\tau$ are very flat, so that slight changes in the input photometry can change the best fitting values for $\tau$ very easily. 
Finally, when combining all the models, the best fit is provided by the SSP model with age of 1.5\,Myr within [0.1,211]\,Myr, and best mass of $\rm2.2\cdot10^8\,M_{\odot}$ within [0.6,4.8]$10^8\,M_{\odot}$. The metallicities and the dust contents of all the cases remain the same as in the previous results.  Thus the inclusion of IRAC upper limits for ID3 leads to best fitting parameters which are in agreement with the previous ones for all the cases, and to slightly smaller $2\sigma$ intervals for the age and mass. \\

\begin{figure*}
\centering
\includegraphics[width=1.2\columnwidth]{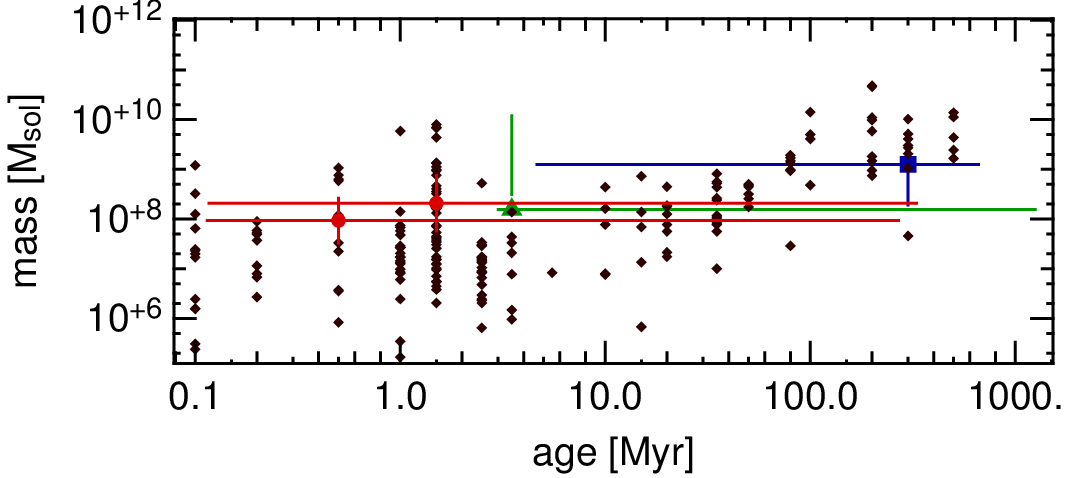}
\caption{\small: Age and mass results for SED fitting runs with all SFHs combined on the objects from the literature and the system discussed in this paper. The plot shows the objects of R11 (green triangle), Z12 (blue square), B13 (black diamonds), and ID2\&3 (red points).
Error bars show the intervals in mass and age within which the total probability reads $95.45\,\%$ (corresponding to a $2\sigma$ confidence interval).
To improve on the clarity of the figure we did not plot the intervals for the \citet{Bradley2013} objects.}
\label{fig:massage}
\end{figure*}

\begin{figure*}
 \centering
 \includegraphics[width=1.2\columnwidth]{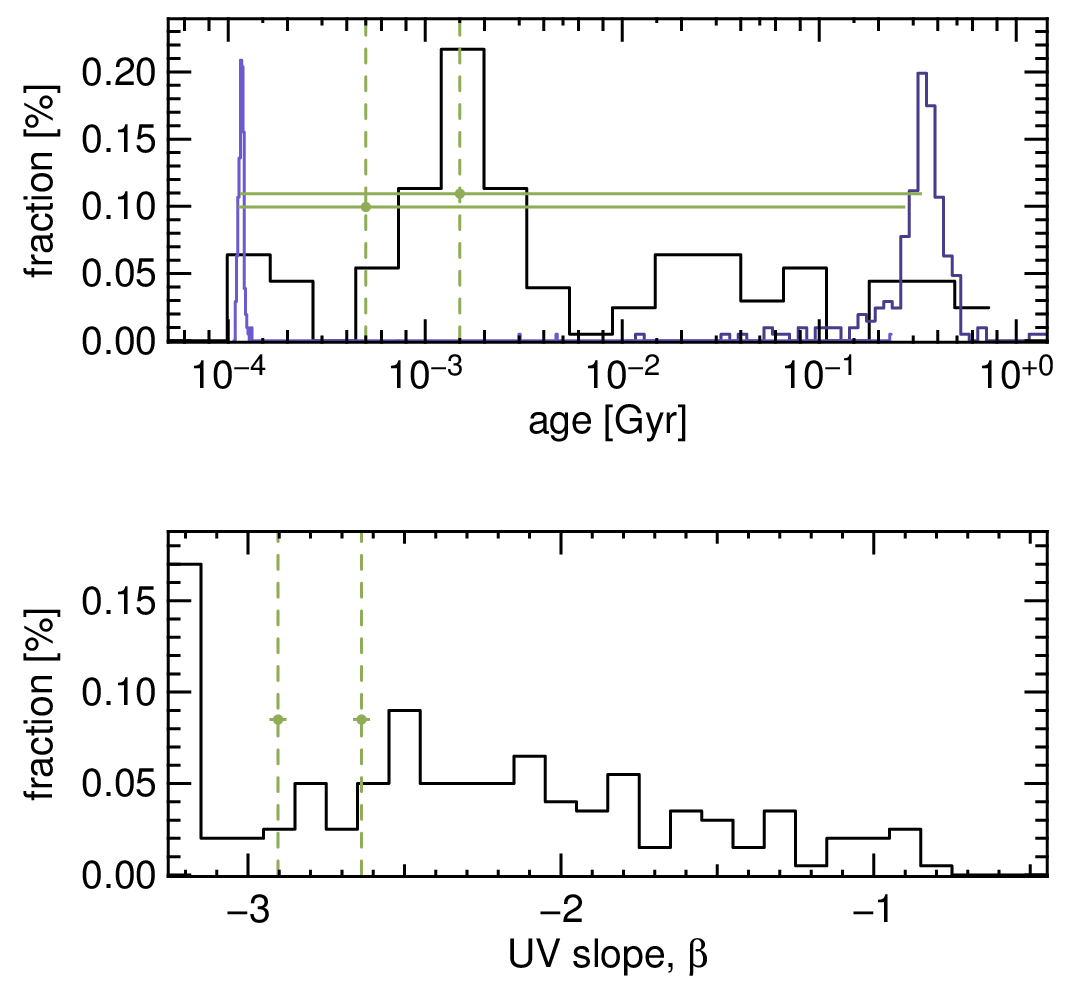}
 \caption{\small \textit{Upper panel}: Histogram of the best ages for the $z\sim6$ candidates from R11, Z12, and B13 (in black). We plot in light and dark blue the age lower and upper limits respectively for the $2\sigma$ confidence level intervals of the $z\sim6$ candidates. In green we plot ID2\&3. The $2\sigma$ confidence level intervals are too large to well constrain the ages of the sources, however our multiple lensed system surely belongs to the young sample of $z\sim6$ galaxies. \textit{Lower panel:} Histogram of the UV slope $\beta$ as measured from the best fitting SED templates for the B13 candidates (in black) and for our \hiz galaxy (in green).}
\label{fig:agelimits}
 \end{figure*}

To robustly compare our results with the literature, we perform the same fitting procedure (using CSPs and SSPs models) on other known $z\sim6$ lensed sources selected in the field of the CLASH clusters \citep[i.e.,][]{Richard2011,Zitrin2012,Bradley2013}, for which we have the same photometry as for our candidate.
The SED fitting performed in our work uses similar model parameters as \citet[][hereafter Z12]{Zitrin2012}.
The main difference is in the lowest model ages, which is $0.1\,\rm{Myr}$ in our work and $5\,\rm{Myr}$ in Z12.
For the $z\sim6.2$ quadruply lensed galaxy in the field of MACS0329, we get results that are consistent with Z12, for a low mass ($\rm M \sim10^9\,M_{\odot}$) young  galaxy. Our best age (300\,Myr) is slightly higher, but still consistent within the $2\sigma$ confidence level. \\
\citet[][hereafter R11]{Richard2011}, unlike us, adopt in their SED fitting models a Sapleter IMF and smaller ranges for the metallicity ($Z$ within $[0.2,1]\,Z_{\odot}$) and ages (within 10\,Myr and 1\,Gyr).
Our best age is much younger than the age range predicted in R11, although our $2\sigma$ confidence level age interval of $\sim[3,10^3]$\,Myr covers the age range of [640,940] Myr given in R11.
As a consequence, also the stellar mass is found to be quite different, given that our best value is much smaller than the mass estimated by R11 ( although they are anyhow consistent within our $2\sigma$ confidence level mass interval).
In table~\ref{tab:SEDfittinglit} we summarise our results for these two lensed systems, and provide the physical properties estimated in the reference works.\\ 
We also perform the SED fitting for the 208 galaxy candidates at $z\sim6$ from \citet[][hereafter B13]{Bradley2013}. 
We plot in Fig.~\ref{fig:massage} the best fitting ages versus masses for these candidates, together with the results we get for our system, R11 and Z12. Our candidates have age and mass similar to many young candidates from B13. In Fig.\ref{fig:agelimits} we show the histograms for best age, with the lower- and upper- age limits, for B13 and our candidates. Although the large $2\sigma$ confidence level intervals, we see that our multiple lensed system is definitely among the young sample of $z\sim6$ galaxies. Moreover we also estimate the UV slope $\beta$ from the best SED fitting templates for all the sources and we get that our candidate belongs to the sample of galaxies with very steep UV slope ($\beta\sim2.6-2.9$).
Moreover, although the large $2\sigma$ confidence level intervals on the ages, our  SED fitting suggest a very young best age for our system,  compared to the other $z\sim6$ multiply lensed galaxies (Z12 and R11), while the Z12 and R11 have confirmed older ages, in agreements with the results from literature.
   
\begin{table*}
\caption{Best fitting parameters for the objects of R11 and Z12.
Columns are the same as in Tab.~\ref{tab:SEDfitting}.}
\centering
\footnotesize
\begin{tabular}{r|c|c|c|c|c|c|l}
\hline
\hline
ID &  $\tau$ / Gyr & $Z\ /\ Z_\odot$ & age [age] / Myr & $M_{*}\ [M_{*}]\ /\ 10^9\,M_{\odot}$  &  $\beta_{UV}$ & $A_{\mathrm{V}}$  & Reference\\
\hline
MACS0329 &  0.1 & 0.005 &300   [4.7,700] & 1.3 [0.2,1.5] & $-2.32\pm0.02$ &  0.0    & this work\\
                        & -- & 0.5  & 180           & 1  & $-2.50\pm0.06$ &  --  &  Z12\\
\hline
A383 &   SSP & 0.005 & 3.5  $[2.9,1.3\cdot10^3]$ & 0.2  [0.3,11.4] &$-2.03\pm0.02$ &  0.8        & this work\\
                          &   0.5 &--     &    [640,940]                     & $6.3^{+2.8}_{-1.2}$            &   $-2$        &  --  &R11\\

\hline  
\end{tabular}

\label{tab:SEDfittinglit}
\end{table*}

\section{Summary and Conclusions}
\label{sec:conclusions}
We report the discovery of a young quintuply lensed galaxy candidate at z$\sim6$ in the field of the galaxy cluster RXC J2248  ($\rm z_{cl}=0.348$).
We identify this system as four i-dropouts, plus a central lensed image which we detect once we remove the BCG from the dataset images. 
The lensed images have colours consistent within the errors, which are in agreement with the colour prediction for starbust galaxies at $z>5$. Moreover, they have photometric redshift of $z\sim6$, with a well defined peak in the PDF(z) and no photo-z solution at low-z.
We perform the strong lensing analysis of the cluster, using 11 systems of multiple images as constraints, and we find that our model predicts the $\rm z\sim6$ multiple lensed system with an accuracy of $0.8\arcsec$. 
The magnifications predicted for the lensed images are between 2.2 and 8.3.
Referring to the two images with the best photometry, we estimate the delensed UV luminosity to be $ L_{\rm{UV}}\sim0.5L^*$ at $z=6$. From the observed NIR colours, we estimate the UV slope of our source to be $\beta=-2.89\pm0.38$. 
We perform a SED fitting of the observed photometry, using SSPs and CSPs SED templates, and derive that our candidate has  subsolar metallicity ($Z/Z_\odot<0.2$), low dust content $(A_V\sim0.2-0.4)$,  stellar mass M$\sim10^8M_{\odot}$ and best age of $\sim1\rm Myr$. Although the age is not well constrained, we can set an upper limit of $\sim300\rm Myr$, given the $2\sigma$ uncertainties that we get. We verify that including the shallow IRAC photometric upper limits in the SED fitting leads to similar results, with slightly smaller $2\sigma$ intervals on the mass and age. 
We finally compare our predicted ages and masses, with the $z\sim6$ candidates selected up to date in the field of all CLASH clusters, and we find that our multiply lensed galaxy has a young age and low mass, similar to some of the objects from B13. However, compared to the other known multiply lensed $z\sim6$ galaxies, our SED fitting suggests a younger age for our multiple lensed system. This source adds to the several multiple lensed objects known at \hiz, characterized for being five times lensed, with a central image identified in the very inner region of the lensing cluster.\\

After the submission of the paper to the journal, three multiple images of this system (ID2,3,4) have been spectroscopically confirmed at z=6.11 by \citealt{Balestra13} in a dedicated VIMOS Large Programme to
follow-up CLASH high-z objects. An independent redshift measurement with VLT/FORS2 is also reported in \citealt{Boone13}, who initially associated this system with a bright sub-mm LABOCA source. Such an association is not supported by the spectroscopic properties described in \citealt{Balestra13} and the hard UV slope reported in this paper.

\section*{Acknowledgements}
This work is supported by the Transregional Collaborative Research Centre TRR 33 - The
Dark Universe and the DFG cluster of excellence ``Origin and Structure of the Universe". 
The CLASH Multi-Cycle Treasury Program (GO-12065) is based on observations made with the NASA/ESA Hubble Space Telescope. The Space Telescope Science Institute is operated by the Association of Universities for Research in Astronomy, Inc. under NASA contract NAS 5-26555. The Dark Cosmology Centre is funded by the DNRF. %AZ was partly supported by contract research ``Internationale Spitzenforschung II/2-6'' of the Baden W\"rttemberg Stiftung.
Support for AZ is provided by NASA through Hubble Fellowship grant HST-HF-51334.01-A awarded by STScI. 
%Part of this work
He was also partly supported by contract research ``Internationale Spitzenforschung II/2-6'' of the Baden W\"urttemberg Stiftung.

\addcontentsline{toc}{chapter}{Bibliography}
\bibliographystyle{mn2efix}
\bibliography{z6_lensed_gal_rxj2248}

\appendix
\section{SED Fitting Results}
\label{app:sedfit}

\begin{figure*}
\centering
\leavevmode
\columnwidth=.48\textwidth
\vspace{0.2cm}
\includegraphics[width=\columnwidth]{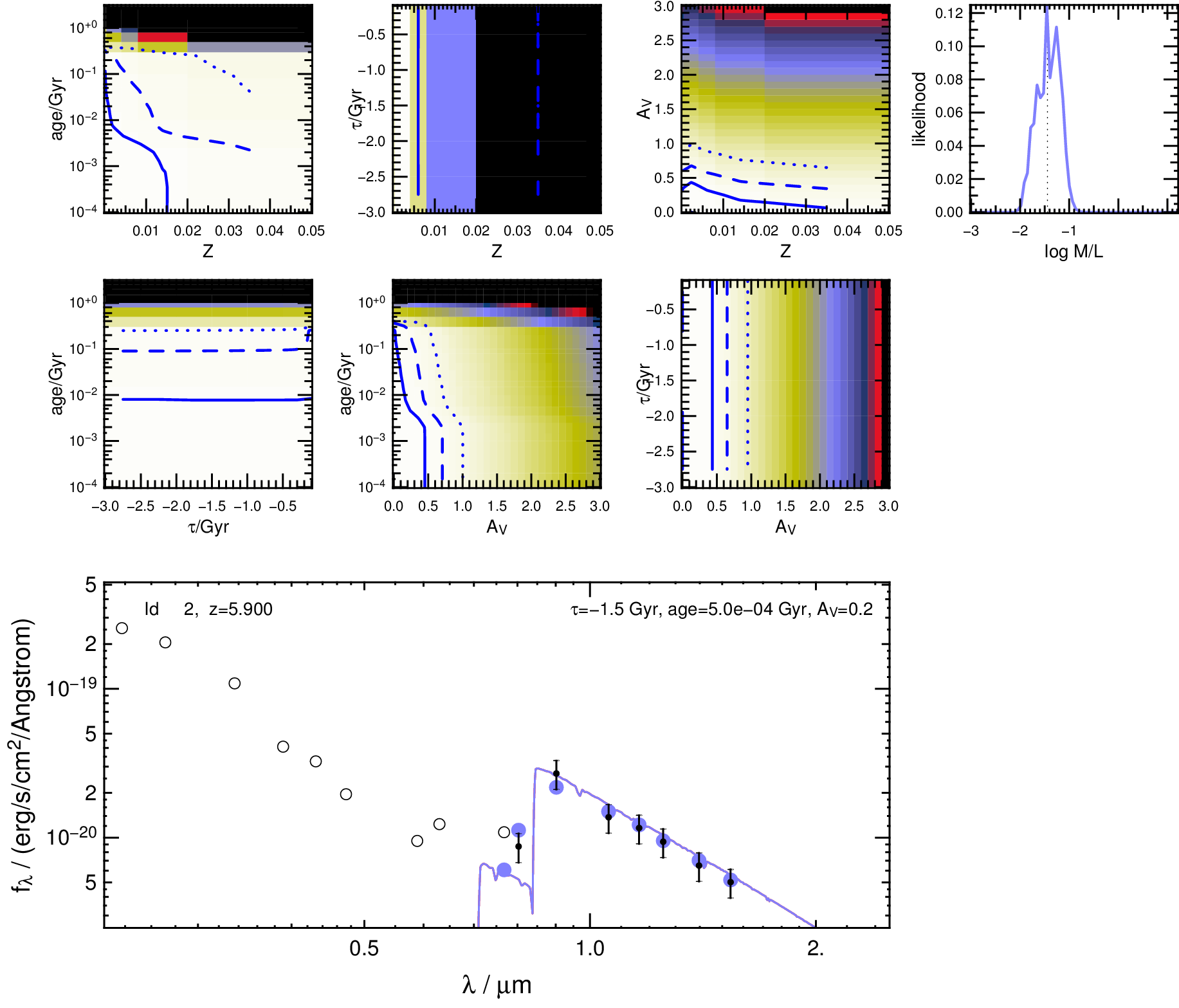}
\includegraphics[width=\columnwidth]{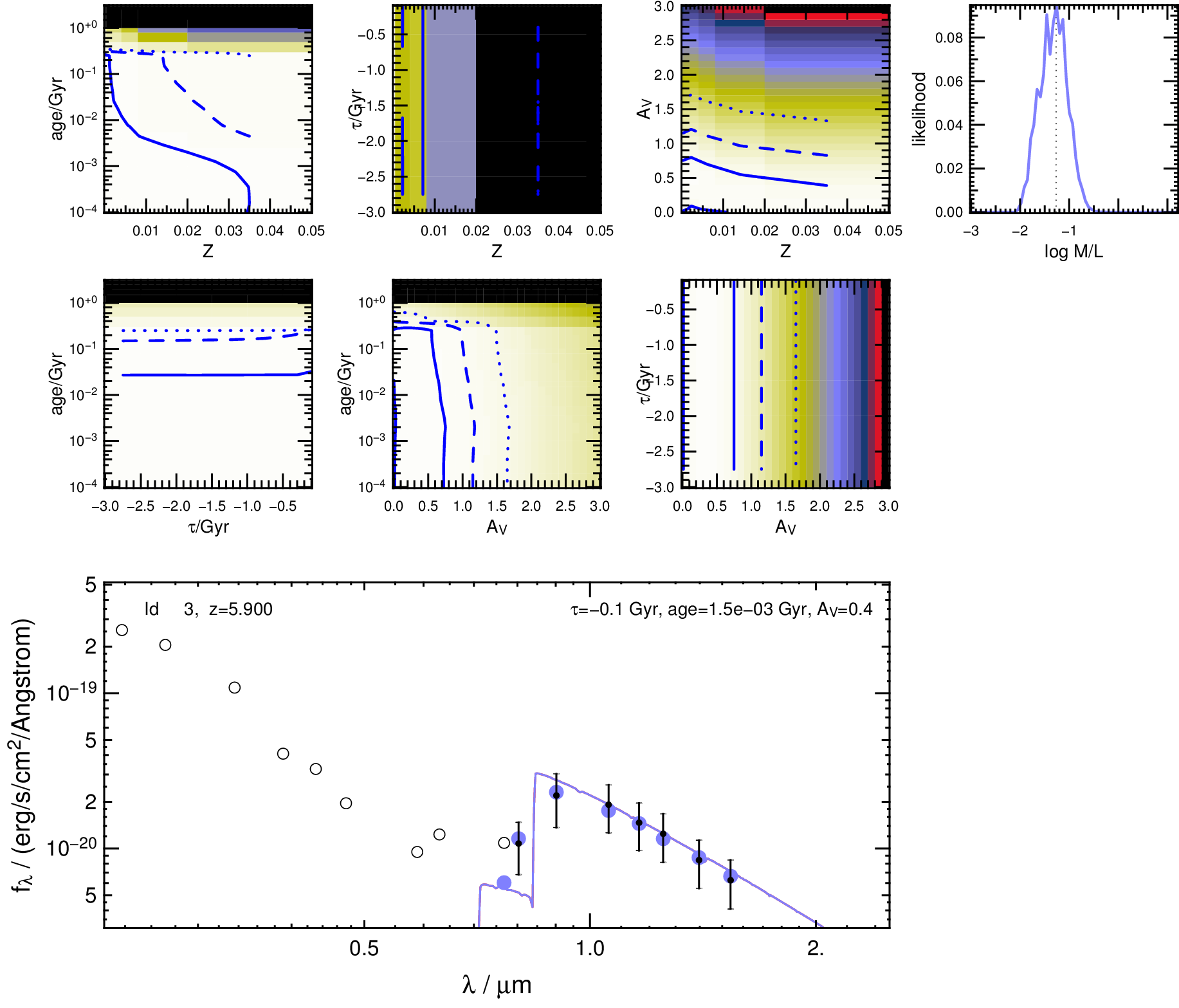}
\caption{\small: SED fitting results for ID2\&3 and models with negative $\tau$ SFRs.
The lower panel shows the input photometry and errors with black points.
Empty circles denote the fluxes in the dropout filters where the fluxes are considered upper limits.
The best fitting model SED is shown in blue and the convolved fluxes in the detection bands are displayed by filled circles.
The density plots in the upper panels show the likelihood distributions of the SED fit in two-dimensional parameter spaces.
Blue lines denote the $1\sigma$ (dotted), $2\sigma$ (dashed), and $3\sigma$ (solid) confidence levels.
Finally, the likelihood distribution of the mass-to-light ratio in the $V$ band is plotted in the upper right panel.
The dotted line in this panel denotes the $M/L$ ratio of the best fitting model.
}
\label{fig:incSFH}
\end{figure*}

\begin{figure*}
\centering
\leavevmode
\columnwidth=.48\textwidth
\vspace{0.2cm}
\includegraphics[width=\columnwidth]{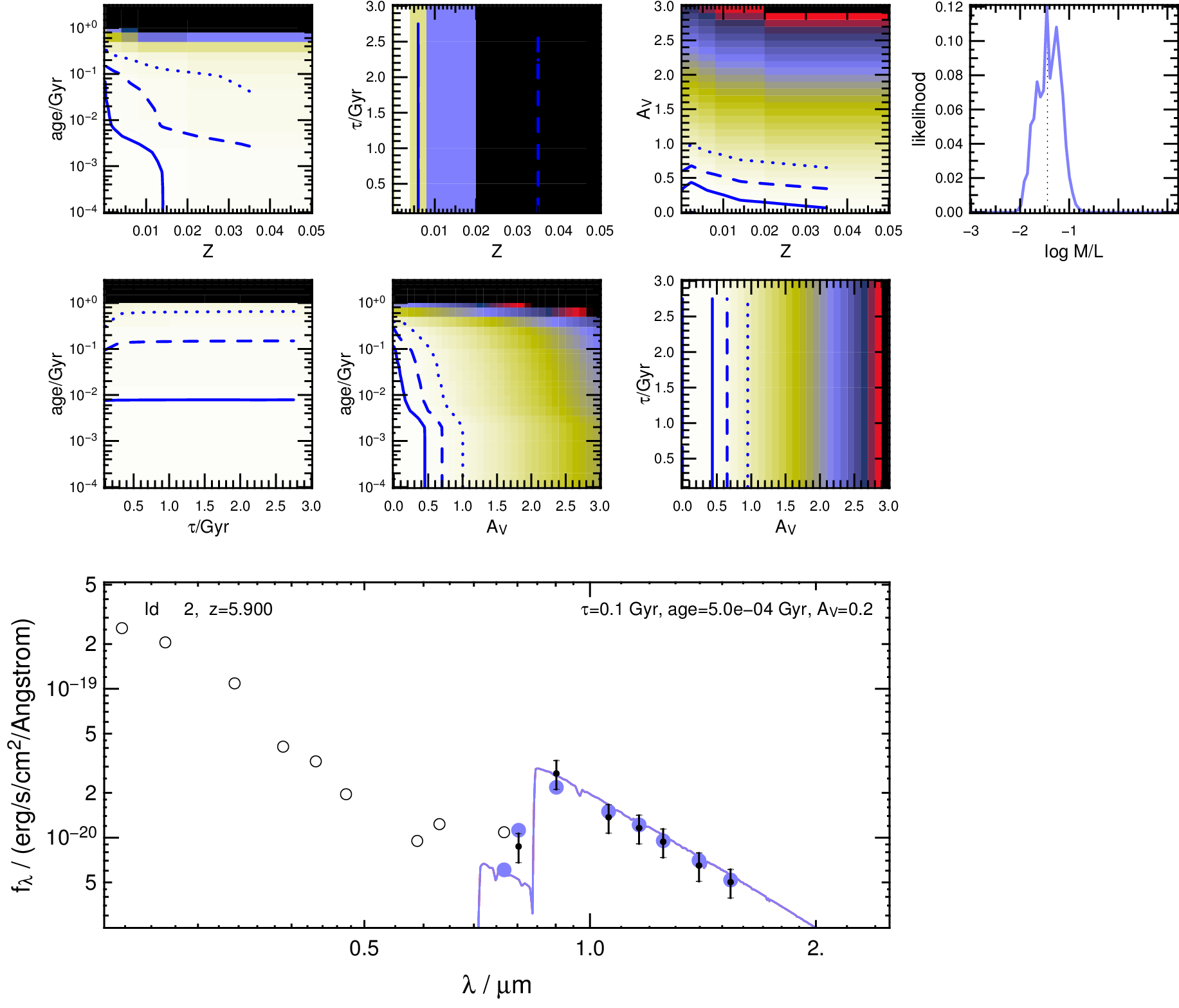}
\includegraphics[width=\columnwidth]{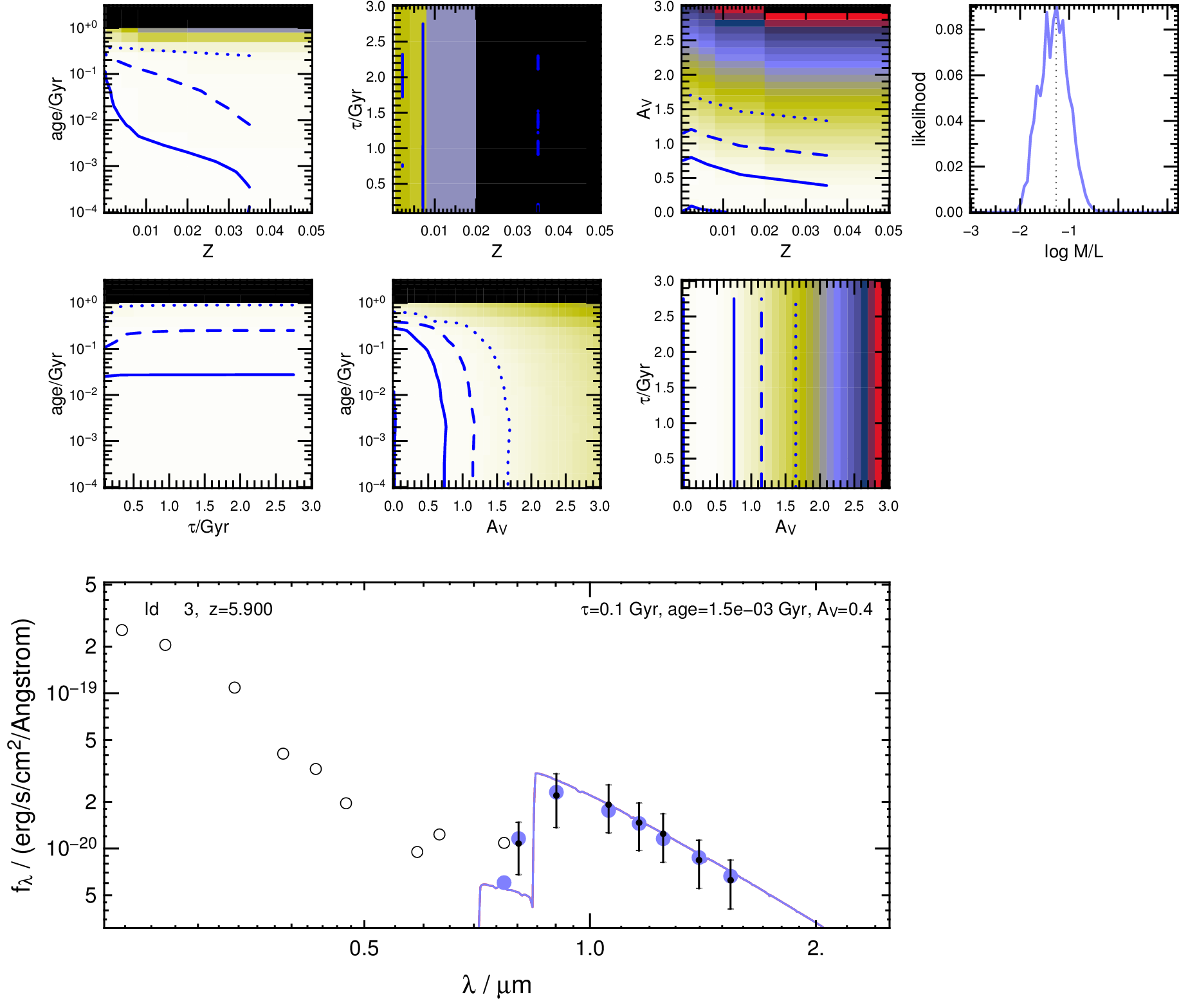}
\caption{\small: SED fitting results for ID2\&3 and models with positive $\tau$ SFRs.
For a detailed description on the plot see Fig.~\ref{fig:incSFH}.}
\label{fig:decSFH}
\end{figure*}

\begin{figure*}
\centering
\leavevmode
\columnwidth=.48\textwidth
\vspace{0.2cm}
\includegraphics[width=\columnwidth]{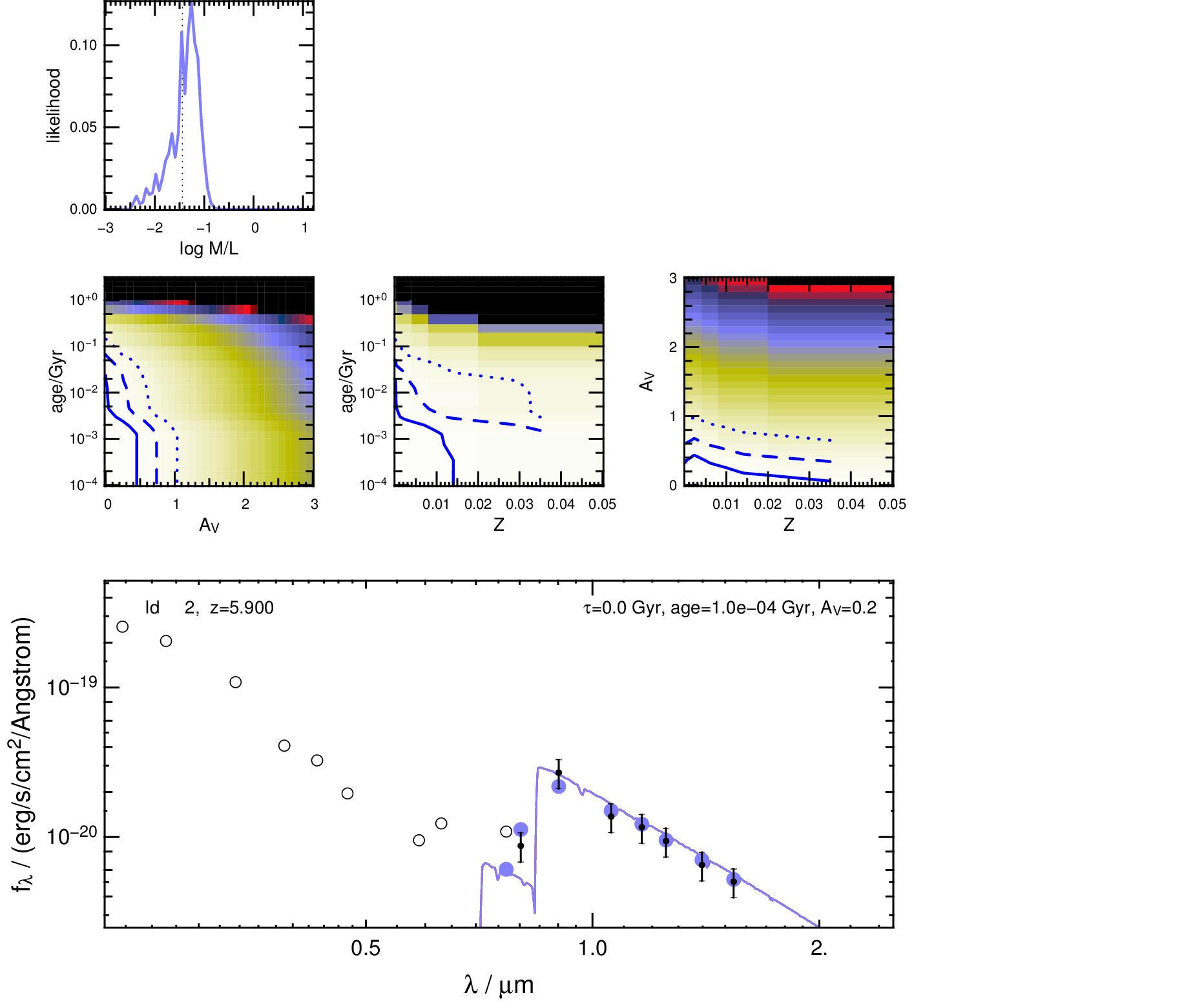}
\includegraphics[width=\columnwidth]{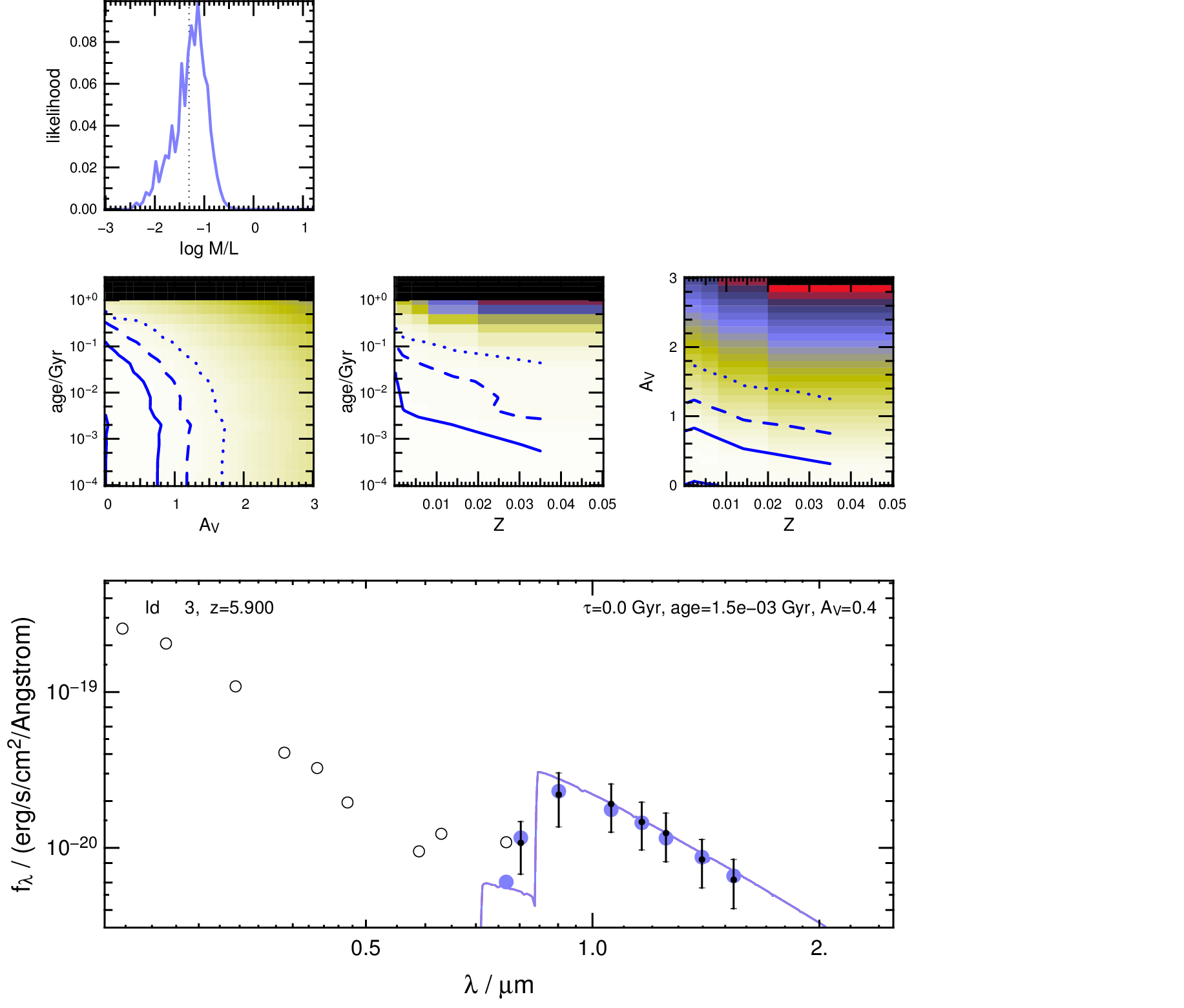}
\caption{\small: SED fitting results for ID2\&3 and SSP models.
For a detailed description on the plot see Fig.~\ref{fig:incSFH}.}
\label{fig:sspSFH}
\end{figure*}

\begin{figure*}
\centering
\leavevmode
\columnwidth=.48\textwidth
\vspace{0.2cm}
\includegraphics[width=\columnwidth]{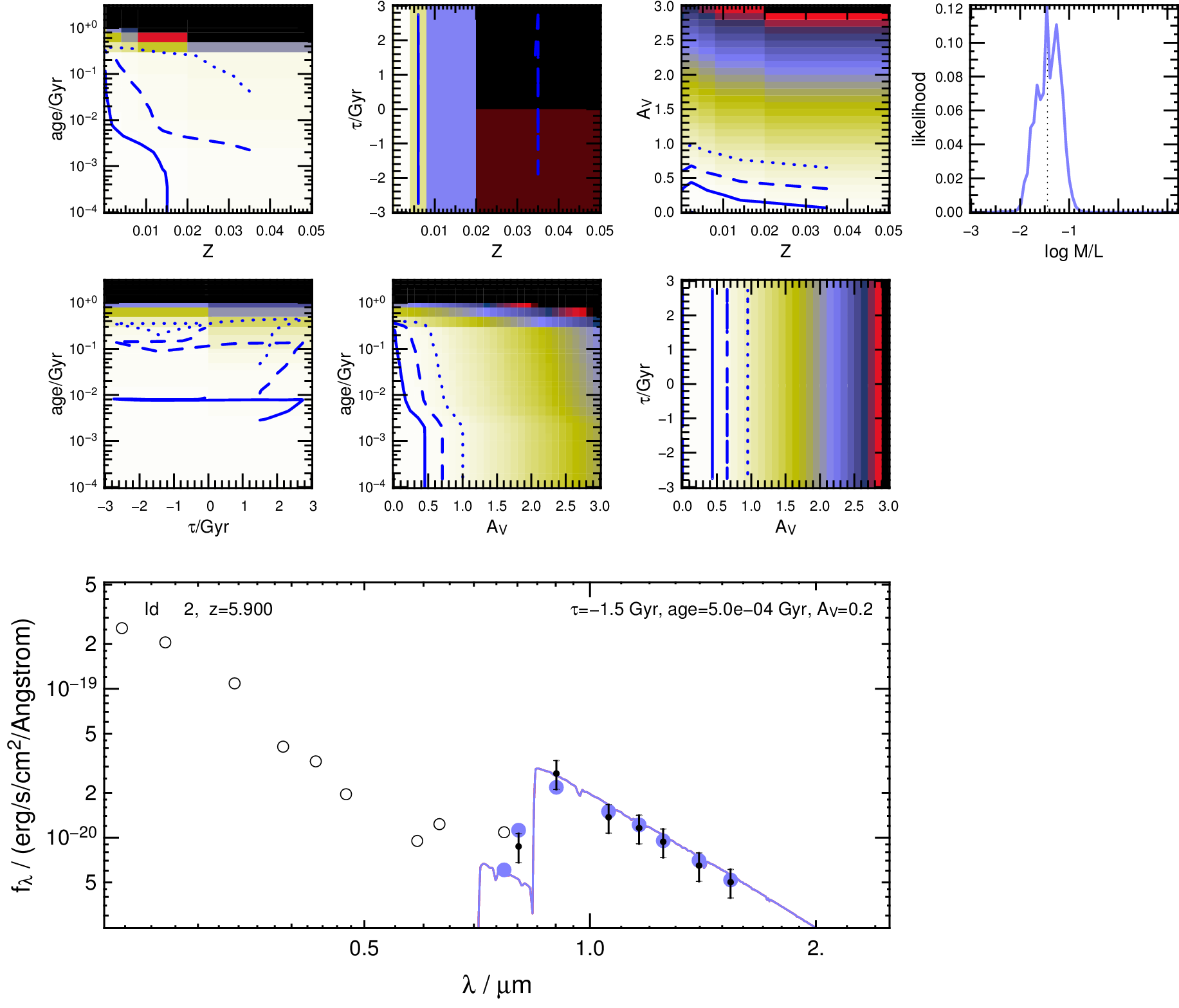}
\includegraphics[width=\columnwidth]{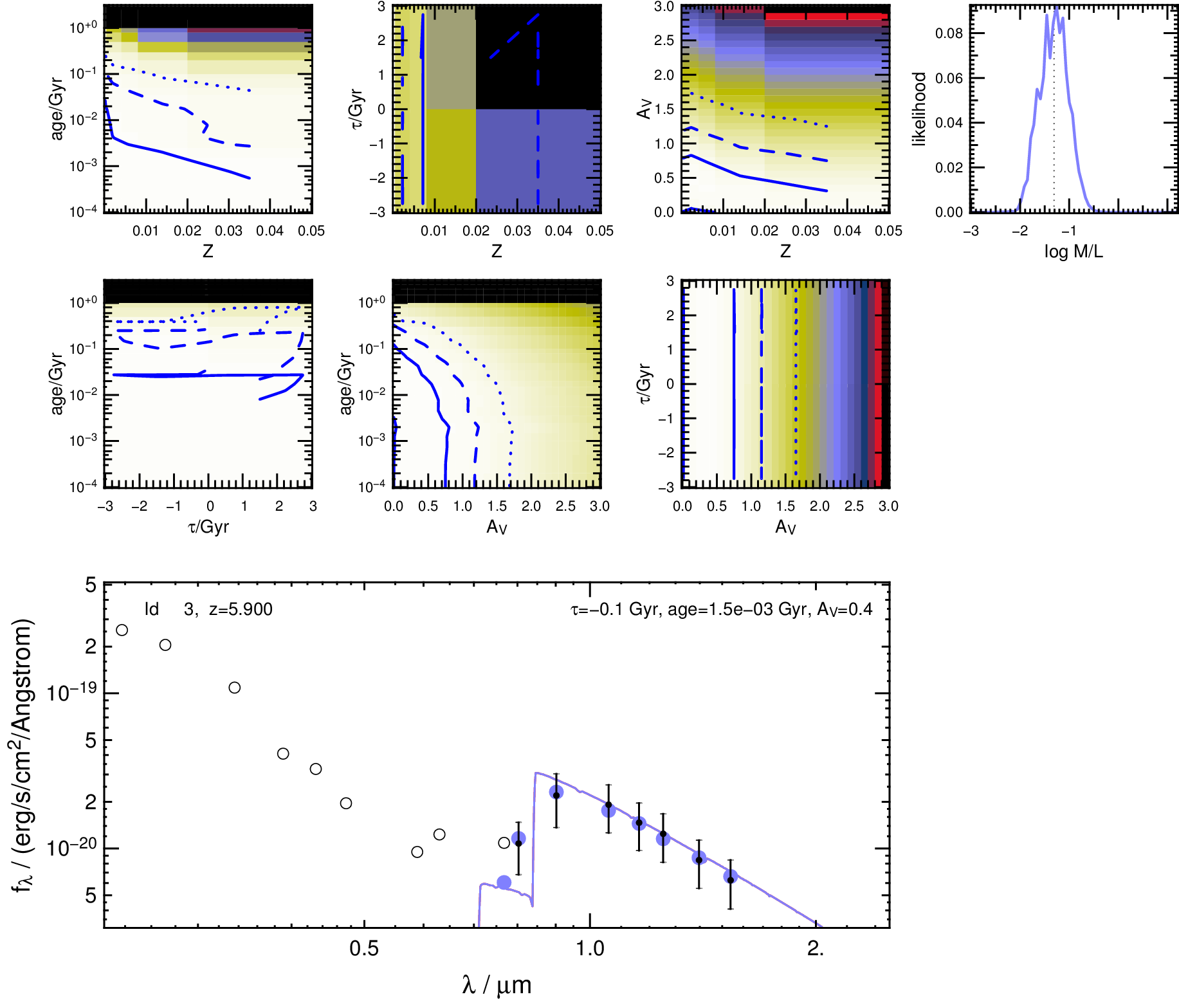}
\caption{\small: SED fitting results for ID2\&3 and all models (CSPs and SSPs).
For a detailed description on the plot see Fig.~\ref{fig:incSFH}.}
\label{fig:allSFH}
\end{figure*}

\begin{figure*}
\centering
\leavevmode
\columnwidth=.55\textwidth
\vspace{0.2cm}
\includegraphics[width=\columnwidth]{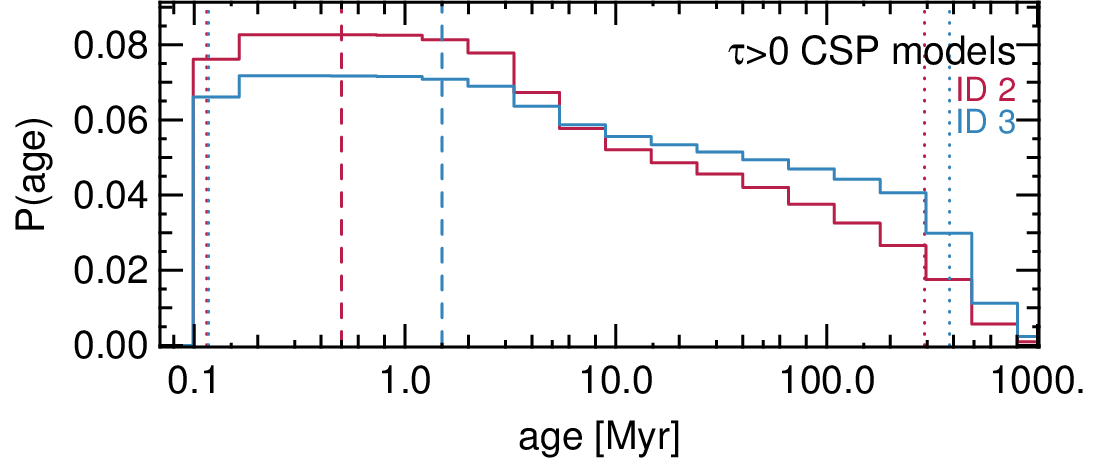}
\caption{\small: PDF of the model age for ID2\&3, marginalized over the other fitting parameters for a fit with $\tau>0$ SFR models.
The dashed lines are the ages of the best fitting models.
Between the dotted lines the total probability is $95.45\,\%$ (corresponding to a $2\sigma$ confidence interval).}
\label{fig:PageincSFH}
\end{figure*}

\begin{figure*}
\centering
\leavevmode
\columnwidth=.55\textwidth
\vspace{0.2cm}
\includegraphics[width=\columnwidth]{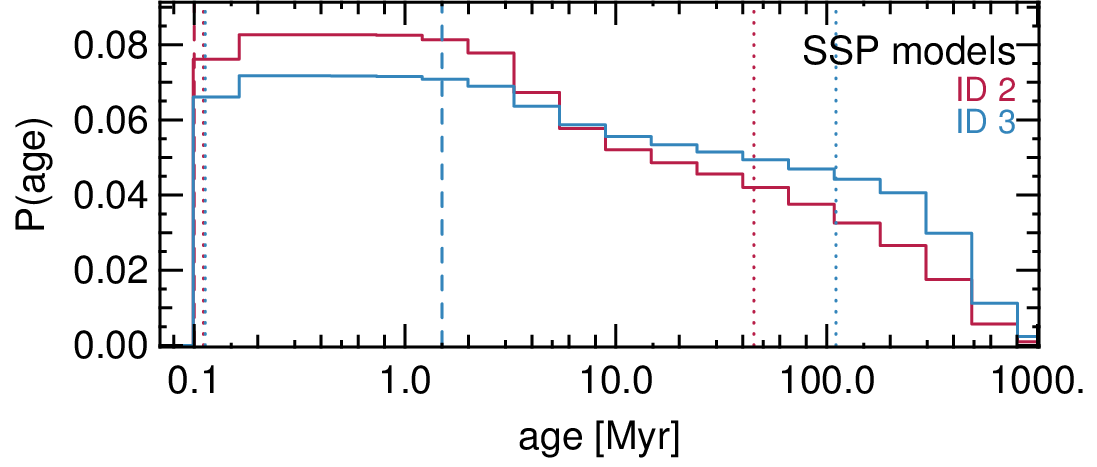}
\caption{\small: PDF of the model age for ID2\&3, marginalized over the other fitting parameters for a fit with SSPs only.
The dashed lines are the ages of the best fitting models.
Between the dotted lines the total probability is $95.45\,\%$ (corresponding to a $2\sigma$ confidence interval).
The best fitting age does not necessarily lie within the interval, as is the case for ID2.}
\label{fig:PagesspSFH}
\end{figure*}

\begin{figure*}
\centering
\leavevmode
\columnwidth=.55\textwidth
\vspace{0.2cm}
\includegraphics[width=\columnwidth]{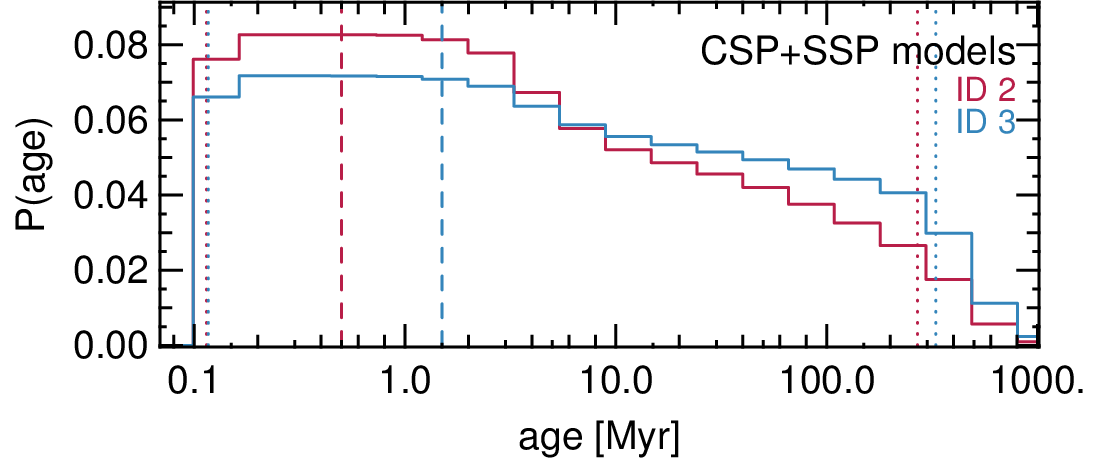}
\caption{\small: PDF of the model age for ID2\&3, marginalized over the other fitting parameters where the model set comprises all created model SEDs (SSPs and CSPs).
The dashed lines are the ages of the best fitting models.
Between the dotted lines the total probability is $95.45\,\%$ (corresponding to a $2\sigma$ confidence interval).}
\label{fig:PageallSFH}
\end{figure*}

We show here the results of the SED fitting from Sec.~\ref{sec:Properties} of the multiple lensed system in greater detail.
We concentrate again only on ID2\&3 which have the cleanest photometry.
Figs.~\ref{fig:incSFH} and \ref{fig:decSFH} display the results for exponentially increasing ($\tau<0$) and decreasing ($\tau>0$) SFRs.
Fig.~\ref{fig:sspSFH} shows the fitting results with SSP models only, whereas Fig.~\ref{fig:allSFH} displays the results when we perform the SED fit to all models (SSPs and CSPs) combined.
The lower panels of Figs.~\ref{fig:incSFH} to \ref{fig:allSFH} display the best fitting SED and the photometry.
The panels in the middle and upper rows show the 2-dimensional likelihood distributions of fitting parameters ($Z$, $\tau$, model age, $A_\mathrm{V}$), whereas the probability distributions of the mass-to-light ratios ($M/L$) in the $V$ band are shown in the upper panels.
The blue contours in the likelihood distributions outline the $1\sigma$- (solid), $2\sigma$- (dashed), and $3\sigma$ (dotted) confidence levels.
The filter bands in which the S/N ratio does not exceed 1 (the dropout filters f225w to f775w) are considered upper limits in the SED fit (lower panels).\\
Figs.~\ref{fig:PageincSFH} to \ref{fig:PageallSFH} show furthermore the probability distributions of the model ages, marginalized over the other fitting parameters.
The upper limits of the $2\sigma$ interval in age is smallest for the fit with SSP models (Fig.~\ref{fig:PagesspSFH}).
This is because all stars are assumed to form at the time of formation and therefore the mean stellar age of a SSP is higher than that of a CSP with the same formation redshift observed at the same cosmic time.\\

\begin{figure*}
\centering
\leavevmode
\columnwidth=.55\textwidth
\vspace{0.2cm}
\includegraphics[width=\columnwidth]{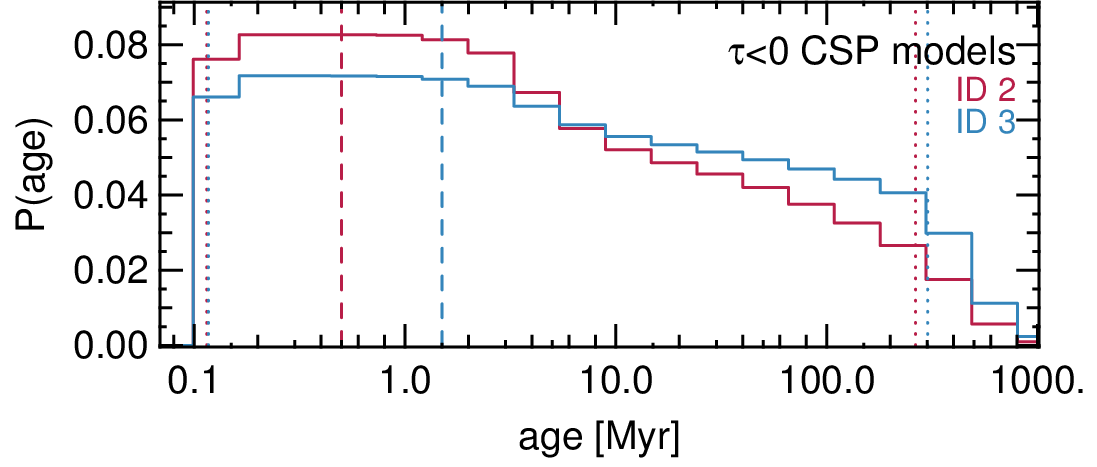}
\caption{\small: PDF of the model age for ID2\&3, marginalized over the other fitting parameters for a fit with $\tau<0$ SFR models.
The dashed lines are the ages of the best fitting models.
Between the dotted lines the total probability is $95.45\,\%$ (corresponding to a $2\sigma$ confidence interval).}
\label{fig:PagedecSFH}
\end{figure*}

\bsp

\label{lastpage}

\end{document}